\documentclass[floatfix, twocolumn,pre, footinbib]{revtex4}
\usepackage{psfrag}
\usepackage{amsmath,amsfonts,amssymb, epsfig}
\usepackage{latexsym}
\usepackage{graphicx}
\usepackage{subfigure}

\newcommand{\avg}[1]{\left\langle #1\right\rangle}

\newcommand{\depar}[2]{\frac{\partial#1}{\partial#2}}

\newcommand{\C}{\mathcal{C}}

\begin{document}

\title{Driven lattice gas of dimers coupled to a bulk reservoir}

\author{Paolo Pierobon} 
\affiliation{Arnold Sommerfeld Center and CeNS, Department of Physics, Ludwig-Maximilians-Universit\"at
M\"unchen, Theresienstrasse 37, D-80333 M\"unchen, Germany}
\affiliation{Hahn-Meitner Institut, Abteilung Theorie, Glienicker Str.100, D-14109 Berlin, Germany}

\author {Thomas Franosch}
\affiliation{Arnold Sommerfeld Center and CeNS, Department of Physics,
Ludwig-Maximilians-Universit\"at M\"unchen,  Theresienstrasse 37, D-80333 M\"unchen, Germany}
\affiliation{Hahn-Meitner Institut, Abteilung Theorie, Glienicker
Str.100, D-14109 Berlin, Germany}

\author {Erwin Frey} 
\affiliation{Arnold Sommerfeld Center and CeNS, Department of Physics,
Ludwig-Maximilians-Universit\"at M\"unchen,  Theresienstrasse 37, D-80333 M\"unchen, Germany}

\date{\today}

\begin{abstract}
  We investigate the non-equilibrium steady state of a one-dimensional
  (1D) lattice gas of dimers. The dynamics is described by a totally
  asymmetric exclusion process (TASEP) supplemented by attachment and
  detachment processes, mimicking chemical equilibrium of the 1D
  driven transport with the bulk reservoir. The steady-state phase
  diagram, current and density profiles are calculated using both a
  refined mean-field theory and extensive stochastic simulations. As a
  consequence of the on-off kinetics, a new phase coexistence region
  arises intervening between low and high density phases such that the
  discontinuous transition line of the TASEP splits into two
  continuous ones. The results of the mean-field theory and
  simulations are found to coincide. We show that the physical picture
  obtained in the corresponding lattice gas model with monomers is
  robust, in the sense that the phase diagram changes quantitatively,
  but the topology remains unaltered. The mechanism for phase
  separation is identified as generic for a wide class of driven 1D
  lattice gases.
\end{abstract}

\maketitle 
\section{Introduction}

Driven lattice gas models have recently received much attention due to
their possible application to intracellular transport \cite{foot1}.
The connection of non-equilibrium transport to biology has been
recognized almost 40 years ago by modelling ribosome motion on mRNA.
In two pioneering papers MacDonald et al.\
\cite{macdonald-gibbs-pipkin:68, macdonald-gibbs:69} have introduced a
one-dimensional (1D) driven lattice gas model which later became known
as the Totally Asymmetric Simple Exclusion Process (TASEP).

In this model each particle occupies a site on a one-dimensional
lattice and advances stochastically uni-directionally (hence the term
``totally asymmetric''). The non-trivial properties arise by imposing
the constraint that each site may at most be occupied by one particle,
and that moves are forbidden if the target site is already occupied
(hence the term ``exclusion''). Of particular interest are systems with
open boundary conditions where particles may enter and leave the 1D
lattice at the left or right boundary with rates $\alpha$ and $\beta$,
respectively. 

Remarkably, this 1D driven system with open boundaries exhibits a
non-trivial phase diagram with $\alpha$ and $\beta$ as control
parameters \cite{krug:91}.

For low extraction rate $\beta$ a high density (HD) phase occurs,
where the density profile is controlled by the right boundary.
Symmetrically, for low injection rate $\alpha$ a low density (LD) phase
appears, where the density profile is determined by the left boundary.
For high values of both $\alpha$ and $\beta$ the bulk density is
independent of the boundary conditions and the system carries its
maximal current (MC).  

Exact solutions were found both for periodic and open boundary
conditions through exact methods, such as recursive relations, Bethe
and matrix product ansatz
\cite{derrida-domany-mukamel:92,schuetz-domany:93}.

The stochastic fluctuations have been investigated in terms of static
and dynamic correlation functions by exact methods
\cite{derrida-etal:93,schuetz:93,degier-essler:05} and
phenomenological approaches \cite{dudzinsky-schuetz:00,
  pierobon-etal:05}.  The rich phenomenology, as well as the
availability of exact results established TASEP as on of the
paradigmatic models in non-equilibrium statistical mechanics
\cite{b:schuetz:01,derrida:98,b:schmittmann-zia:95}.

Over many years the TASEP has served as a testing ground for a
mathematical analysis of non-equilibrium problems. More recently, the
application to biology reentered the focus of attention in the context
of systems of extended particles. For example, ribosomes bind to mRNA
occupying twelve codons and progressively advance by a single codon
through two cycle of GTP hydrolysis (elongation step). The
implications of mutual steric hindrance of $\ell$-mers (monomers,
dimers, 12-mers) have been studied by stochastic simulations and
`refined mean-field theory' for $\ell$-TASEP \cite{lakatos-chou:03,
  shaw-zia-lee:03}.  The picture already found for monomers has been
corroborated also by exact methods (Bethe ansatz
\cite{alcaraz-bariev:99} and mapping to zero range process
\cite{schoenherr-schuetz:04}).

Intracellular transport, where a molecular motor (kinesin) is moving
along a molecular track (microtubule) is yet another example of 1D
non-equilibrium transport in a biological system (see e.g.\
\cite{b:alberts-etal:02, b:howard:01}). In marked difference to
directed motion of ribosomes on mRNA here there is exchange of
molecular motors with the surrounding cytosol acting as a particle
reservoir.  This observation has motivated a recently proposed
extension of TASEP, by supplementing the unidirectional hopping with
Langmuir (on-off) kinetics
\cite{lipowsky-klumpp-nieuwenhuizen:01,parmeggiani-franosch-frey:03}.
Single motor experiments suggest that both processes compete on the
scale of the microtubule, i.e.\ a motor explores a significant
fraction of the track before detaching \cite{b:howard:motors}. To
capture such an interplay mathematically, a \emph{mesoscopic limit}
has been suggested, where local adsorption-desorption rates have been
rescaled in the limit of large but finite systems, such that the gross
rates are comparable to the injection-extraction rates at the
boundaries \cite{parmeggiani-franosch-frey:03}. The study of this
system has shown an unexpectedly rich and new phase diagram with phase
coexistence regions. Surprisingly in the mesoscopic limit the density
profiles show a sharp discontinuity (shock separating low from high
density phase) or cusp in a large portion of the phase diagram.

Here we propose a model for intracellular transport incorporating both
coupling to a reservoir as well as the finite extension of particles.
The model is motivated by the fact that many processive molecular
motors (e.g.\ kinesins, dyneins and myosin V) are composed of two
heads (trail and lead) that bind specifically each to a subunit of the
molecular track ($\alpha$- and $\beta$-tubulin in case of
microtubules). Hence, we concentrate on particles which occupy two
sites, i.e.\ dimers, and advance by a single site per hopping event.

The theoretical challenge arises from finding an appropriate and
quantitative description that captures both the on-off kinetics as
well as the $\ell$-TASEP features on an equal footing. Whereas for
monomers a straightforward decoupling of correlations yields a
consistent mean-field theory, such an approach fails even to describe
the chemical equilibrium of a pure on-off kinetics
\cite{lakatos-chou:03}. In the following paragraphs we review some
aspects of our model where the geometric constraints usually play an
important role introducing non-trivial static and dynamic
correlations.
 
At the \emph{static} level the geometrical frustration emerges already
in the equilibrium thermodynamics of the one-dimensional \emph{Tonks
  gas}. Identical extended particles are distributed according to the
grand canonical ensemble on a one dimensional lattice. A hard core
repulsion reduces the task to a combinatorial problem
\cite{tonks:36,b:thompson:88}. It turns out that the single site
occupation densities do not determine the configurational
probabilities, contrary to the monomer case. There the \emph{Langmuir
  kinetics} specified in terms of attachment and detachment rates
$\omega_A$ and $\omega_D$, or equivalently the binding constant
$K=\omega_A/\omega_D$. The coverage density is simply determined by a
single site consideration, resulting in the \emph{Langmuir isotherm}:
$\rho_{I}= K/(K+1)=\omega_A/(\omega_A+\omega_D)$. This is to be
contrasted with the dimer case where a full evaluation of the grand
canonical partition function is required leading to an equilibrium
coverage density \cite{mcghee-hippel:74}:
\begin{eqnarray}
  \label{eq:onoff}
  \rho_I=1-\frac{1}{\sqrt{1+4K}}\, .
\end{eqnarray}

At a \emph{dynamical} level a non-trivial approach to the steady state
is found in the on-off kinetics in the fast attachment limit: at short
time scales only deposition processes are frequent, while detachment
processes are still unlikely. The kinetics belong to the class of
problems referred to as \emph{random sequential adsorption (RSA)} and
the so-called ``Flory plateau'' \cite {flory:39} is reached
exponentially fast. After this transient the particles can detach (on
longer time scales) freeing new gaps; this process results in a
long tail relaxation where the isotherm is approached with a power law
$t^{-1/2}$. Such behavior has been explained using the mapping of the
gap dynamics to a 1D reaction diffusion system $A+A\to\emptyset$ (for
a review see \cite{evans:93,mattis:98,b:privman:97} and references
therein) for which this anomalous power law dynamics is known.  This
two step relaxation has been found also in other models like in
irreversible deposition of dimers (i.e.\ RSA) with diffusion
\cite{privman-nielaba:92} and is intrinsically related to the extended
nature of the particles.

Driving a system of extended particles far from equilibrium, e.g.
introducing the $\ell$-TASEP dynamics, suggests a variety of competing
time scales and ordering phenomena, i.e.\ non-equilibrium phase
transitions.  The purpose of this paper is to shed light on density
and current profiles in the stationary state for a model incorporating
such aspects, as well as to provide a complete analytical theory
supported by simulated data.

This paper is organized as follows: in Sec.~\ref{sec:model} we specify
the model under consideration and introduce the used formalism.
Section \ref{sec:mc} exemplifies results of the stochastic
simulations and puts extracted features into proper perspective.  In
Sec.~\ref{sec:mf} we construct a suitable set of mean-field equations
relying on results for $\ell$-TASEP and on-off kinetics.  In
Sec.~\ref{sec:mfpd}, with the help of these equations, we rationalize
the MC data and study the complete phase diagram. In the Conclusion,
Sec.~\ref{sec:conclusion}, we discuss the robustness of the model,
give a possible generalization to $\ell$-mers and summarize our main
results.

\section{Model and notation}
\label{sec:model}
\begin{figure}[htbp]
  \begin{center}
    \psfrag{alpha}{$\alpha$}
    \psfrag{beta}{$\beta$}
    \psfrag{a}{$a$}
    \psfrag{i=1}{$i=1$}
    \psfrag{i=N}{$i=N$}    
    \psfrag{L=1}{$L=1$}
    \psfrag{tau=1}{$\tau=1$}
    \psfrag{omega_A}[][c]{$\omega_A$}
    \psfrag{omega_D}[][c]{$\omega_D$}
    \includegraphics[width=\columnwidth]{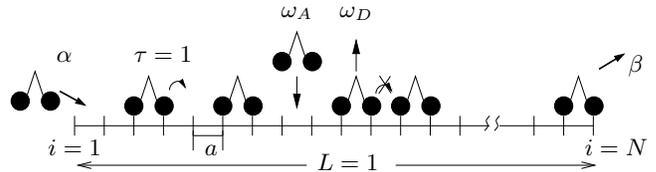}
    \caption{ \label{fig:model}Schematic representation of the model
      and allowed moves: forward jump (with rate $\tau=1$), entrance
      at the left boundary (with rate $\alpha$), exit at the right
      boundary (with rate $\beta$), attachment (with rate $\omega_A$),
      and detachment (with rate $\omega_D$) in the bulk.}
    \end{center}
\end{figure}

We consider a finite one dimensional lattice with sites labeled
$i=1,\dots,N$, of unit length $L=1$; consequently the lattice spacing
reads $a=L/N$. The first two sites ($i=1,2$) and the last two sites,
($i=N-1,N$) represent, respectively, the left and right boundary. The
lattice is partially covered by dimers, i.e.\ composite objects
consisting of two monomers rigidly tied together. A dimer occupies two
lattice sites, see Fig.~\ref{fig:model}; we refer to the monomer to
the right of the dimer as the \emph{lead head} and correspondingly to
the left monomer as the \emph{trail head}.

A microscopic state $\C$ of the system consists of a configuration of
non-overlapping indistinguishable dimers on the lattice.
Non-overlapping requires that no two heads (trail or lead) occupy the
same lattice site.
We specify the dynamical evolution by the following set of updating
rules:
\begin{itemize}
\item If the lead head of a dimer occupies a site $i=2,\dots,N-1$ and
  the following site is empty, the dimer advances one step with
  unit rate.
\item At site $i=1$ a dimer enters the lattice with its trail head
  with rate $\alpha$ provided that the first two sites are empty.
\item A dimer with its lead head on site $i=N$ leaves the lattice with
  rate $\beta$ emptying the last two sites.
\item Everywhere in the bulk (i.e.\ trail head on a site $i=2,\dots,N-2$) a
  dimer leaves the lattice emptying two sites with a site independent
  detachment rate $\omega_D$.
\item Everywhere in the bulk ($i=2,\dots,N-2$) a dimer enters the lattice
  with its trail head, provided that the considered site as well as
  its right neighbor is empty, with a site-independent attachment rate
  $\omega_A$.
\end{itemize}
The first three rules encode the usual Totally Asymmetric Simple
Exclusion Process of extended objects ($\ell$-TASEP), while the last
two implement the coupling of the lattice to a reservoir of dimers
with fixed chemical potential \cite{foot2}.
  
The rules entail that a complete description of the underlying
stochastic Markov process is given in terms of the time-dependent
probability $P(\C,t)$. The time evolution is governed by the
associated master equation 
\begin{eqnarray}
\label{eq:master}
\partial_t P(\C)=\sum_{\C'\neq \C}\left[W_{\C'\to \C}P(\C',t)-W_{\C\to
    \C'}P(\C,t)\right]\, ,
\end{eqnarray}
where the transition rates $W_{\C'\to \C}$ can be inferred form the
dynamical rules.

It is expected that the system evolves towards a stationary ensemble
$P^{st}(\C)$ for long times. Due to the lack of detailed balance this
macroscopically stationary state does not correspond to an equilibrium
ensemble, i.e.\ there is no Gibbs-Boltzmann measure to be inferred from
thermodynamics arguments.  The macroscopic quantities we aim to
compute are understood as averages over this distribution.  

To exploit probabilistic methods we label sites according to: empty
(state $s=0$), occupied with the trail head (state $s=1$) or with the
lead head (state $s=2$).  The configuration $\C$ can be represented as
a string of occupation number $\C=\{n_1,\dots,n_N\}$, with
$n_i=\{0,1,2\}$. The quantity of fundamental interest in the following
is the average site-dependent dimer density in the stationary state,
In particular we shall derive equations for the density of the lead
head $(n_i=2)$ at site $i$, $\rho_i\equiv\avg{\delta_{n_i,2}}$. Since the
dimers are rigidly connected, the corresponding density for the trail
head at site $i$ simply reads $\rho_{i+1}$.

The probability of having site $i$ in state $s$ will be denoted by
$p(i,s)\equiv P(n_i=s)$. Since the states mutually exclude, the
probability $p(i,2)$ immediately yields the average lead head density:
$\rho_i=p(i,2)$. The coverage density $\rho^c_i$ is sum of the lead
head density and the trail head density, i.e.\ $\rho_i^c=p(i,2)+p(i+1,2)$. Consistently with the notation the joint
probability will read $p(i,s;j,s')$. We also shall need the
conditional probability $p(i+1,s|i,s')$, i.e.\  the probability for
site $i+1$ to be in state $s$ provided site $i$ is in state $s'$.

In the simulations we will concentrate on the average coverage density
$\rho^c_i$, and the spatially resolved current $j_i$, defined as the
flux of particle through site $i$ per unit time. Only averages on the
stationary state will be considered: a comparison between ensemble and
time-moving averages corroborates the hypothesis that the system is
ergodic.

As indicated in the Introduction, a non-trivial competition between
boundary induced phenomena and bulk dynamics arises in the
\emph{mesoscopic limit}, i.e.\ the gross on/off rates, $\Omega_A$ and
$\Omega_D$, should be of the same order of magnitude as the boundary
entry/exit rates $\alpha$ and $\beta$. This requires that the single
site on/off rates, $\omega_A$, $\omega_D$, scale with the system size
$N$ as:
\begin{eqnarray}
  \label{eq:rates}
  \omega_A=\frac{\Omega_A}{N}\textrm{, \hspace{1cm}}\omega_D=\frac{\Omega_D}{N}
\end{eqnarray}
with $\Omega_A$ and $\Omega_D$ fixed.

\section{MC results and description of the phase diagram in general}
\label{sec:mc}
In this section we present the method used to perform numerical
simulations and the results obtained from them. Density profiles much
different from the TASEP motivate our interest in the model.

\subsection{Method}
We have simulated the stochastic equation through a kinetic Monte
Carlo algorithm, to determine the average density profile in the
stationary state with high accuracy.  The results have been obtained
using a random sequential updating algorithm by Bortz, Kalos and
Lebowitz (BKL-method or $n$-fold way)~\cite{bortz-kalos-lebowitz:75,
  b:landau-binder:00}: a list of all sites which are possible
candidates for a successful move is stored and dynamically updated.
The method is (for the present case) faster than the conventional
algorithms and constitutes a reliable way to simulate \emph{real time}
dynamics, although here we focus on equal-time averages in the
stationary state (note that the algorithm is equivalent to
Gillespie's, commonly used in chemical reactions \cite{gillespie:77}).

Explicitly, we keep three lists of sites for the respective allowed
moves: there are $N_{J}$ sites where particles can perform a jump
forward, $N_{A}$ sites that may accept a particle from the reservoir,
and $N_{D}$ sites from which a particle can detach. Furthermore we
keep track of the occupation of the boundaries ($n_0$ and $n_N$).  One
of these moves is selected randomly with the appropriate weights: $ 1
\times N_J, \omega_A \times N_A, \omega_D\times N_D, \alpha
\times(1-n_1), \beta\times n_{N-1}$. Then we increment time by an
interval drawn from an exponential distribution with timescale
$T=(\alpha(1-n_0)+\beta n_N+N_{J}+\omega_A N_{A}+\omega_D
N_{D})^{-1}$. The lists are updated correspondingly and the procedure
is iterated for many events. We have started the algorithm from an
empty configuration and after the stationary state has been reached,
the coverage density and the current have been measured. Moving time
averages typically cover a window of $O(10^6)$ time units. We have
investigated finite-size effects by considering lattices varying from
$128$ up to $4096$ sites.  Furthermore, we have checked the ergodicity
by comparing time-moving and ensemble averages. The longest
simulations have been performed approximately a day on a 1GHz
processor.

\subsection{Results}
\begin{figure}[htbp]
  \begin{center}
    \psfrag{r}[][][1.2]{$\rho^c$}
    \psfrag{x}[][][1.2]{$x$}
    \psfrag{j}[][][1.2]{$j$}
    \psfrag{w=0.001}{$\Omega_D=0.001$}
    \psfrag{w=0.01}{$\Omega_D=0.01$}
    \psfrag{w=0.05}{$\Omega_D=0.05$}
    \psfrag{w=0.1}{$\Omega_D=0.1$}
    \psfrag{w=0.2}{$\Omega_D=0.2$}
    \psfrag{w=1}{$\Omega_D=1$} 
    \psfrag{w=10}{$\Omega_D=10$}
    \includegraphics[width=\columnwidth]{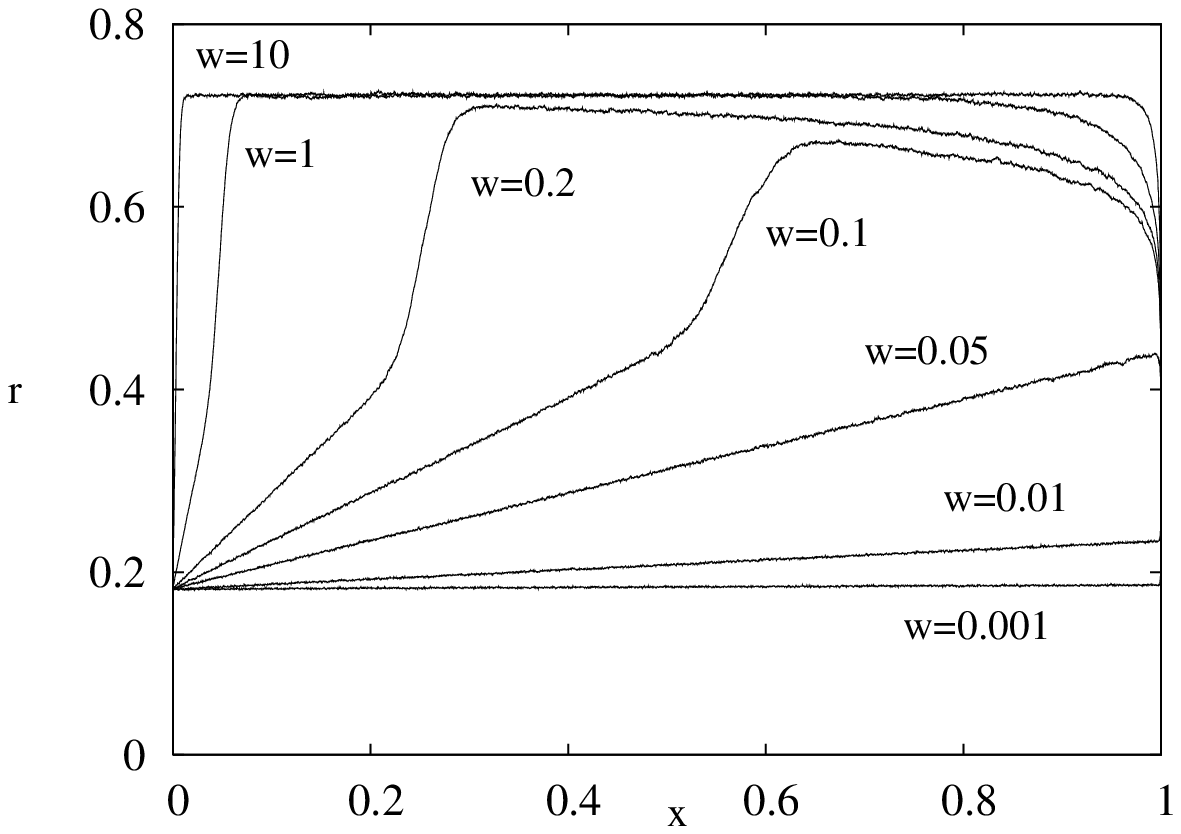}
    \includegraphics[width=\columnwidth]{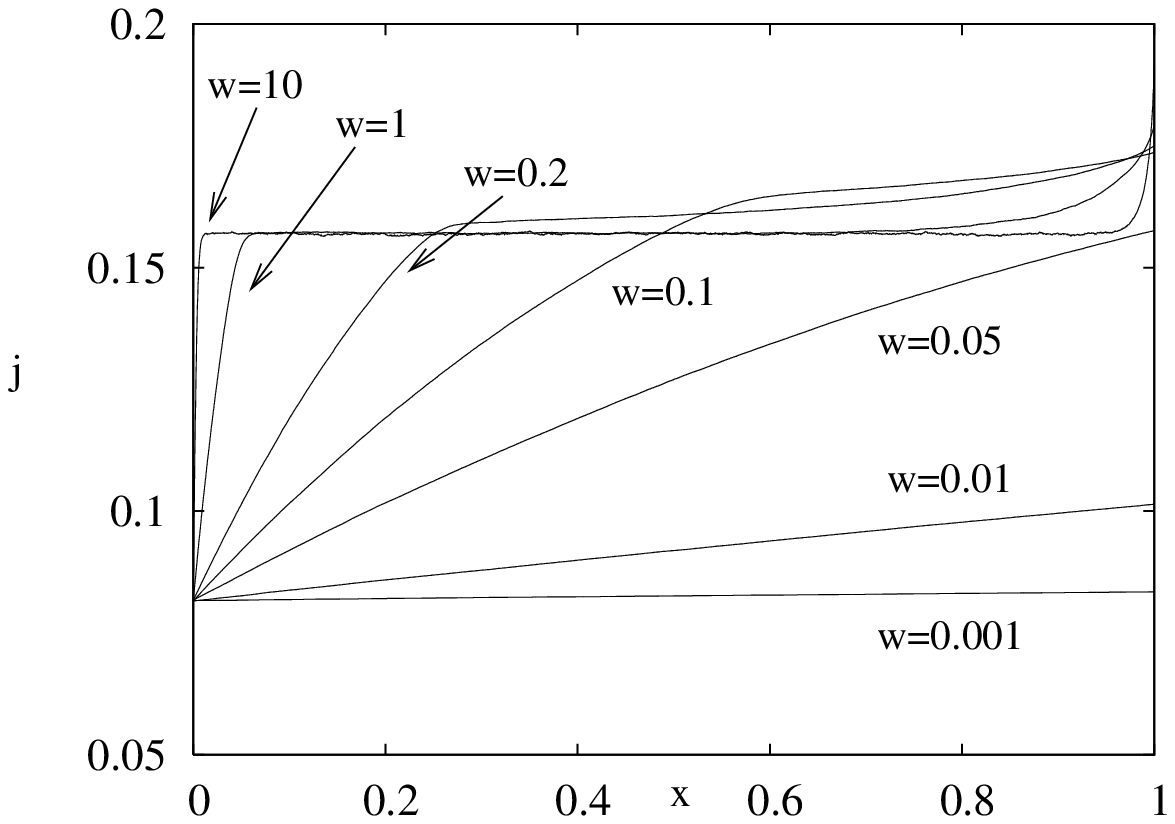}
    \caption{ \label{fig:preliminary}Stochastic simulations results
      for stationary density profiles and corresponding currents for a
      system of $N=4096$ lattice sites.  Parameters have been fixed to
      $\alpha=0.1$, $\beta=0.6$, $K=3$, and $\Omega_D$ are varied as
      indicated in the figures.}
  \end{center}
\end{figure}

We exemplify our results by fixing the binding constant to $K=3$ and
the entrance/exit rates to $\alpha=0.1$ and $\beta=0.6$; see
Fig.~\ref{fig:preliminary}. We have measured the stationary density
profiles as well as the current for different values of the detachment
rate $\Omega_D$. For very small $\Omega_D$ we find almost flat
profiles for both $\rho$ and $j$, in accordance with the picture of
simple dimer TASEP \cite{macdonald-gibbs:69, lakatos-chou:03,
  shaw-zia-lee:03}. On the other hand, for very large detachment rates
the on/off kinetics dominates the bulk of the density profile.
However, contrary to equilibrium thermodynamics of on/off kinetics, a
large directed current is simultaneously present.

For intermediate values of the detachment rate, neither the density
profile nor the current is constant anymore. There appear rather
narrow regions where the density steeply rises from an almost linear
profile to one that approaches the isotherm value.  Such findings are
similar to the ones of monomeric TASEP coupled to Langmuir kinetics
\cite{parmeggiani-franosch-frey:03, parmeggiani-franosch-frey:04},
where it has been shown that the steep increase
corresponds to a domain wall of a coexistence phase in the mesoscopic
limit.

\section{Construction of the MF equation}
\label{sec:mf}
Since the stationary density profiles for TASEP of monomers coupled to
Langmuir kinetics have been successfully described in terms of
mean-field theory, it appears promising to look for an analytical
description for dimers also. However, it turns out that in the case of
dimers the conventional mean-field approach, i.e.\ a decoupling of
correlations at lowest order starting from the quantum Hamiltonian
formulation, see e.g.\ \cite{parmeggiani-franosch-frey:04}, is
inappropriate: the non-trivial correlations, arising from the fact
that the particles are extended, manifest themselves already at the
level of pure TASEP of dimers \cite{lakatos-chou:03}.

The failure of conventional mean-field approach requires the
introduction of phenomenological methods. The aim of this section is
to identify useful strategies that are capable to deal first with the
simple on/off kinetics of extended objects without directed motion and
second with pure TASEP for dimers. We will then combine the two cases
to obtain a formulation for the complete problem, and we will provide
arguments that the aspects of correlations are captured on a reliable
level for both processes.

The density profile in the mesoscopic limit will be described in terms
of a balance equation. The evolution of the course-grained dimer
density is determined by: (a) a flux $j$ of dimers within the lattice
that encodes the asymmetric diffusion, (b) two source terms that
represent the on-off kinetics of dimers:
\begin{eqnarray}
  \label{eq:general}
  \partial_t\rho=-\partial_x j+F_A-F_D\, .
\end{eqnarray}
The non-trivial task is to provide reasonable expressions for the
current and the sources in terms of the local density.  In
Sec.~\ref{sec:mfpd} we shall compare the analytical results to
stochastic simulations and demonstrate the validity of this approach.

\subsection{ON-OFF kinetics of dimers}
\subsubsection{Probabilistic approach}
\label{subsec:prob}
The dynamics of the on-off kinetics in a rate equation approach is described 
by the difference between
a gain term $F_i^A$ and a loss term $F_i^D$:
\begin{eqnarray}
  \label{eq:flux1}
  \partial_t \rho_i=F_i^A-F_i^D\, .
\end{eqnarray}
These two terms are proportional to the attachment and
detachment rates, $\omega_A$ and $\omega_D$. Clearly the detachment
flux is exactly given by $F_i^D=\omega_D \rho_i$.\cite{foot3}

An attachment event at site $i$ (lead head) requires both sites $i$
and $i-1$ to be empty (i.e.\ in state $(0)$). The corresponding
attachment flux then reads $F_i^A=\omega_Ap(i-1,0;i,0)$, and knowledge
of the joint probability is needed. In order to obtain closed
equations for the lead head density we break the hierarchy of joint
probabilities resulting from the master equation. Here, we rely on an
{\em ad hoc} approximation for the conditional probabilities:
$p(i,0|i-1,0)\simeq p(i,0|(i-1,0)\lor (i-1,2))$.\cite{foot4} The idea
is that the probability of having site $i$ empty is independent of
having a lead head or a hole in the site $i-1$: note that this is
precisely what usually the mean-field approximation does in monomeric
lattice gases, where correlations are broken factorizing the density:
$\avg{n_in_{i+1}}=\avg{n_i}\avg{n_{i+i}}$. In Appendix \ref{app:cp}
we show that this assumption yields a closed expression for the joint
probability:
\begin{eqnarray}
  \label{eq:MF1}
  p(i,0;i-1,0)=
   \frac{p(i-1,0)p(i,0)}{1-p(i,2)}\, .
\end{eqnarray}
Eliminating the probabilities for single empty sites and collecting
results, the rate equation for the lead head density reads:
\begin{eqnarray}
  \label{eq:LK} 
  \partial_t \rho_i=\omega_A \frac
  {(1-\rho_{i-1}-\rho_{i})(1-\rho_{i}-\rho_{i+1})}{1-\rho_{i}}-\omega_D \rho_i\, .
\end{eqnarray}

As a next step we perform a continuum limit. Relabelling position as
fraction of the system size, $x\equiv i/N$, the density $\rho_i$
becomes a function $\rho(x)$ in the mesoscopic limit $N\to \infty$.
The average density at the neighboring sites are obtained by expansion
in powers of the lattice constant $a=1/N$:
\begin{eqnarray}
  \label{eq:contlim}
  \rho(x\pm a)=\rho(x)\pm a\partial_x\rho(x)+\frac 1 2 a^2\partial^2_x\rho(x)+O(a^3)\, .
\end{eqnarray}
Introducing a new rescaled time $\tau= a t$ and eliminating the single
site on/off rates $\omega_A$ and $\omega_D$ in favor of the gross
kinetic rates $\Omega_A$ and $\Omega_D$ we obtain the rate equation
in the mesoscopic limit:
\begin{eqnarray}
  \label{eq:LK2}
  \partial_\tau\rho=\Omega_A\frac {(1-2\rho)^2}{1-\rho}-\Omega_D\rho+O(a^2)\, .
\end{eqnarray}

It is natural to expect a density limited to the range $[0,1/2]$,
because there cannot be more than one lead head per site. In the
stationary regime the physical solution corresponds to the constant
density $\rho_I$, Eq.~(\ref{eq:onoff}), determined solely by the binding
constant $K=\Omega_A/\Omega_D$. This value of the \emph{isotherm} is
consistent with the one obtained by McGhee and Von Hippel
\cite{mcghee-hippel:74} for the general case of $\ell$-mers.

\subsubsection{Tonks gas approach}
To corroborate the results presented above, we compare the isotherm of
the rate equation with the one computed from the grand canonical
partition function of the Tonks gas. In particular, we shall identify
the fugacity $z$ with the binding constant $K$.

The statistical mechanics of the Tonks gas on an open lattice consists
of distributing extended objects on the $N$ sites, respecting the
excluded volume constraint.  In the case of the canonical ensemble
this reduces to the combinatorial problem of counting the number of
ways to distribute $n$ dimers on $N$ lattice sites. Using the standard
trick to represent the two occupied sites of each dimer by a ``stick''
and the $N-2n$ empty sites by ``balls'' \cite{b:thompson:88}, one
immediately concludes
\begin{eqnarray}
  \label{eq:parttonks}
  Z(n, N)=\binom{N-n}{n} \, .
\end{eqnarray}
The corresponding grand canonical partition function is readily
evaluated using the previous result 
\begin{eqnarray}
  \label{eq:GCparttonks}
  \mathcal{Z}_{N}&=&\sum_{n=0}^N z^n Z(n,N)\nonumber\\
  &=& \frac{\left(\sqrt{1+4z}+1\right)^{N+1}+\left(\sqrt{1+4z}-1\right)^{N+1}}{2^{N+1}\sqrt{1+4z}}
\end{eqnarray}
where $z$ denotes the fugacity \cite{foot5}. The average density is
then obtained in the thermodynamic limit by
\begin{eqnarray}
  \label{eq:densfug}
  \rho=\lim_{N\to\infty}\frac{1}{N}z\depar{
}{z} \ln \mathcal{Z}_{N} =\frac 1 2\left(1-\frac{1}{\sqrt{4z+1}}\right) \, .
\end{eqnarray}

Since the master equation of pure on/off kinetics fulfills detailed
balance, the stationary state is given in terms of the grand canonical
ensemble.  By a single detachment event a configuration $\C$ of $n$
dimers connects to some new configurations $\C'$ of $n-1$ dimers.  For
such configurations the detailed balance condition implies
$P(\C)/P(\C') = \omega_A/\omega_D=K$. Conversely, the ratio can be
determined also from the grand canonical Boltzmann factors.  Since,
apart from the total exclusion, energy does not enter the problem, the
probabilities are determined by the number of dimers only:
$P(\C)=z^n/\mathcal{Z}_{N}$ and $P(\C') = z^{n-1}/ \mathcal{Z}_{N}$.
Combining both expressions, we conclude $z=K$ as in the case of
monomers.

The last result shows that the ``isotherm'' obtained using {\em ad
  hoc} approximations (sub-Section \ref{subsec:prob}) is at least
consistent with thermodynamics. Let us mention that in equilibrium the
situation is probably better than in the general case, since
correlations are quickly washed out by the coupling to the reservoir.
The relaxation towards equilibrium can be rather different from the
naive picture of rate equations, see e.g.\ \cite{frey-vilfan:02}.

\subsection{TASEP of dimers}

In pure TASEP particles cannot leave or enter the track except at the
boundaries. Correspondingly in the bulk a conservation law holds
\begin{eqnarray}
  \label{eq:tasep}
  \partial_t \rho_i=j_{i-1}-j_i\, .
\end{eqnarray}
The currents $j_i$ can be determined by probabilistic arguments. A
hopping event of a dimer at site $i$ requires first that site $i$ is
occupied by a lead head and second that site $i+1$ is empty. Since we
fixed the hopping rate to unity this yields
\begin{eqnarray}
  \label{eq:TASEPj}
  j_i=p(i,2;i+1,0)\, .
\end{eqnarray}
Hence, also in the case of pure TASEP the master equation induces a
whole hierarchy of joint probabilities. To close the equation we rely
on the same \emph{ad hoc} approximation for the conditional
probabilities as for the on/off kinetics. In the appendix we show that
this implies for the current a closed expression
\begin{eqnarray}
  \label{eq:flux}
j_i=\frac{\rho_{i}(1-\rho_{i+1}-\rho_{i+2})}{1-\rho_{i+1}}\, .
\end{eqnarray}
Performing the continuum limit as in Eq.~(\ref{eq:LK2}) one obtains
\begin{eqnarray}
  \partial_{\tau}
  \rho=-\partial_x\left[\frac{\rho(1-2\rho)}{1-\rho}+\frac{1-2\rho^2}{(1-\rho)^2}\,\frac a 2\,\partial_x\rho+O(a^2)\right]\,.
\end{eqnarray}
Since, at the very end, we consider large but finite systems we have
kept the leading correction in $a$, which will turn out to be relevant
for the formation of shocks and boundary layers.

The equation of continuity in the bulk has to be supplemented with
appropriate boundary conditions. Relying again on the same \emph{ad
  hoc} assumption for the conditional probability, and performing the
continuum limit, one finds
\begin{subequations}
  \label{eq:bcpp}
  \begin{equation} 
    \label{eq:bcpp1}
  \rho(0)=\frac{\alpha}{1+\alpha}\, ,
  \end{equation}
  \begin{equation}
    \label{eq:bcpp2}
    \rho(1)=\frac{1-\beta}{2}\, .
  \end{equation}
\end{subequations}
The details are presented in Appendix~\ref{app:bc}. The properties of
these equations have been studied in details in
Ref.~\cite{shaw-zia-lee:03}.  The resulting phase diagram is
topologically equivalent to the one of TASEP of monomers. The
continuum analogue for the current-density relation,
Eq.~(\ref{eq:flux}), reads
\begin{eqnarray}
  \label{eq:currden}
 j=\frac{\rho(1-2\rho)}{1-\rho}
\end{eqnarray}
and exhibits a maximum at a distinguished density $\rho^*=1/(2+\sqrt
2)$.  This value will play an important role in the more general case
of TASEP coupled to on/off kinetics.

\subsection{Combining TASEP and on/off kinetics}
As anticipated at the beginning of this section, we identify the
current and the source terms of Eq.~(\ref{eq:general}) and we
re-expressed everything in term of the density.  The current is
described by Eq.~(\ref{eq:currden}) and the on/off kinetics by
Eq.~(\ref{eq:LK2}).  By taking the mesoscopic limit we have introduced
the gross attachment and detachment rates and rescaled the time
consistently in both processes. The balance equation to order $a$ can
be easily rewritten as
\begin{eqnarray}
  \label{eq:MFcont}
  \partial_{\tau}\rho&=&-\partial_x\left[-\frac a 2 \frac{\partial_x\rho}{(1-\rho)^2}+\frac{\rho(1-2\rho)}{1-\rho}\right]\nonumber\\
  &&+\Omega_A\frac{(1-2\rho)^2}{(1-\rho)}-\Omega_D\rho
\end{eqnarray}
with the boundary conditions, Eq.~(\ref{eq:bcpp}). These equations
constitute the mean-field description to be discussed in the following
section.

Equation~(\ref{eq:MFcont}) may also be recast in a form reminiscent of
a non-linear Smoluchowsky equation \cite{b:vankampen:81}:
\begin{eqnarray}
  \label{eq:fokkerplanck}
  \partial_\tau\rho=-\partial_x\left[A(\rho)\rho-B(\rho)\rho\partial_x\rho\right]+C(\rho)
\end{eqnarray}
where the coefficients $A$, $B$ and $C$ depend on $\rho$. The second
coefficient will prove not to be relevant since it scales to zero with
increasing the system size to infinity.

\begin{figure}[htbp]
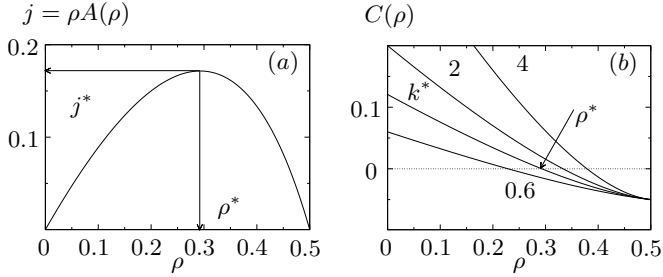

  \begin{center}
    \psfrag{(a)}{$(a)$}
    \psfrag{(b)}{$(b)$}
    \psfrag{r}[t][b]{\vspace{5mm}$\rho$}
    \psfrag{j}{$j=\rho A(\rho)$}
    \psfrag{C}{$C(\rho)$}    
    \psfrag{j*}[tl]{$j^*$}
    \psfrag{r*}[tl]{$\rho^*$}
    \psfrag{0}[]{0}
    \psfrag{0.1}[]{0.1}
    \psfrag{0.2}[]{0.2}
    \psfrag{0.3}[]{0.3}
    \psfrag{0.4}[]{0.4}    
    \psfrag{0.5}[]{0.5}    
    \psfrag{2}{$2$}
    \psfrag{k*}{$k^*$}
    \psfrag{0.6}{$0.6$}
    \psfrag{4}{$4$}
    \begin{tabular}{lr}
      \hspace{0mm}\includegraphics[width=0.45\columnwidth]{ab1.eps}&
      \hspace{5mm}\includegraphics[width=0.45\columnwidth]{ab2.eps}
    \end{tabular}
    \caption{\label{fig:ac} (a) Graph of the current-density relation
      $\rho A(\rho)$ in Eq.~(\ref{eq:fokkerplanck}). (b) Graph of the
      source-drain $C(\rho)$ term in the Smoluchowsky
      equation~(\ref{eq:fokkerplanck}) for different values of $K$ and
      $\Omega_D=0.1$.}
  \end{center}
\end{figure}

We may now interpret $j(\rho)=\rho A(\rho)$ as an effective particle
current and $C(\rho)$ as a source/drain term; see Fig.~\ref{fig:ac}
for an illustration. As will become clear in the following section the
maximum of $j(\rho)$ and he zeros of $C(\rho)$ play an important role
for the analytical from of the stationary solution.

\section{Solution of the mean-field equation and phase diagram}
\label{sec:mfpd}
This section is devoted to a discussion of the phase diagram emerging
from combining TASEP of dimers with on/off kinetics. We shall derive
density profiles within a mean-field approach, identify different
phases and phase boundaries as the kinetic rates are varied. In
particular, we demonstrate the role of the binding constant in the
topological changes of the phase diagram as a certain critical value
is reached.

\subsection{Emergence of shocks and boundary layers}
The mean-field equation for the stationary density profile,
Eq.~(\ref{eq:MFcont}) supplemented with $\partial_{\tau}\rho\equiv 0$, is a
second order differential equation as long as the lattice constant $a$
is small but finite. In the limit $a\searrow0$ the differential equation
degenerates into a first order one.  However, the density profile has
to match the two boundaries, and it appears that the problem is
overdetermined.  Indeed one can construct separately a density profile
fulfilling either the left or right boundary condition. The two
branches will be referred to as $\rho_\alpha$ and $\rho_\beta$,
respectively.  In general, the two branches do not join smoothly
together. One expects the solution of the second order differential
equation to be well approximated by the two branches $\rho_\alpha$ and
$\rho_\beta$ except for a region where the solution rapidly crosses
over from one branch to the other. Upon decreasing the lattice
constant $a$ the crossover region shrinks leading eventually to a
discontinuity of the density profile for $a=0$. Depending on the
position of such a discontinuity $x_{w}$ we refer to it as a
\emph{shock} ($0<x_w<1$) or a \emph{boundary layer} ($x_w=0$ or
$x_w=1$).

To locate the position of the shock or boundary layer we rely on a
very general argument already used in
\cite{parmeggiani-franosch-frey:03, parmeggiani-franosch-frey:04}. The
second order differential equation for the stationary density profile
can be written in the form of a balance equation $\partial_x
j=F_A-F_D$,  where the current along the track reads
\begin{eqnarray}
j=-\frac a 2
\frac{\partial_x\rho}{(1-\rho)^2}+\frac{\rho(1-2\rho)}{1-\rho}\, .
\end{eqnarray}
Integrating  over a small region of width $2\delta x$ around the
shock one obtains
\begin{eqnarray}
  \label{eq:shock1}
  j(x_w+\delta x)-j(x_w-\delta x)=\int_{x_w-\delta x}^{x_w+\delta x}(F_A-F_D)dx\, .
\end{eqnarray}
In the limit $a\searrow 0$ the l.h.s. reduces to $j_\alpha(x_w+\delta
x)-j_\beta(x_w-\delta x)$ (where we have defined
$j_\alpha=\rho_\alpha(1-2\rho_\alpha)/(1-\rho_\alpha)$ the current set
by the left boundary and similarly for the right one $j_\beta$).
Performing the limit $\delta x\searrow 0$ the r.h.s. of
Eq.~(\ref{eq:shock1}) vanishes and yields the matching rule in terms
of the currents
\begin{equation}
j_\alpha(x_w)=j_\beta(x_w)\, .
\end{equation}
Equivalently the rule implies for the densities at the matching point 
\begin{equation}
\label{eq:condition}
\rho_\alpha(x_w)=\frac 1 2 \, \frac{1-2\rho_\beta(x_w)}{1-\rho_\beta(x_w)}\, .
\end{equation}
Let us make a comment. The fact that the in-track current is
continuous at the shock is consistent with the idea of the mesoscopic
limit. The fluxes due to attachment and detachment are important
only on the length scale of the system size. Locally, i.e.\ on the
scale of the lattice constant, the balance equation is dominated
entirely by the unidirectional hopping process (TASEP).

\subsection{Analytical solution}
The left and right branches of the stationary density profile are
determined by solving Eq.~(\ref{eq:MFcont}), once we set the l.h.s.
to zero and discard the second order derivative, i.e.
\begin{eqnarray}
  \label{eq:MFcont2}
  \frac {1}{\Omega_D}\partial_x\left[\frac{\rho(1-2\rho)}{1-\rho}\right]=K\frac{(1-2\rho)^2}{(1-\rho)}-\rho\, ,
\end{eqnarray}
(where the binding constant $K=\Omega_A/\Omega_D$ has been introduced)
which has to be supplemented by the appropriate boundary conditions
Eq.~(\ref{eq:bcpp}). By separation of variables the general solution
$G(\rho)=x+const$ is obtained after a straightforward integration with
\begin{eqnarray}
  \label{eq:integrate}
  G(\rho)&=&\frac{1}{\Omega_DK}\big[\ln(1-\rho)+A_+\ln\left(\rho_{+}-\rho\right)\nonumber\\
&&  +A_-\ln|\rho_{-}-\rho|\big]\, ,
\end{eqnarray}
where we introduced the zeros of the source term
\begin{subequations}
  \begin{equation}
    \rho_\pm\equiv\frac 1 2\left(1\pm\frac{1}{\sqrt{1+4K}}\right)  
  \end{equation}
and the amplitudes
  \begin{equation}
    A_\pm\equiv \frac {K}{1+4K}\frac{1-4\rho_\pm+2\rho_\pm^2}{(1-\rho_\pm)(\rho_\pm-\rho_\mp)}\, .
  \end{equation}
\end{subequations}
The function $G(\rho)$ exhibits a singularity in the physical regime
$0\leq \rho \leq 1/2$ at the isotherm of the on/off kinetics $\rho_I =
\rho_-$. One easily checks that $G(\rho)$ exhibits an extremum when
the corresponding current, Eq.~(\ref{eq:currden}), is maximal, i.e.\ at
$\rho=\rho^*$.  For $\rho_I > \rho^*$ ($\rho_I < \rho^*$) the extremum
of $G(\rho)$ corresponds to a maximum (minimum). A change of topology
of the graph of the solution occurs once both distinguished densities
coincide $\rho_I = \rho^*$, which happens at a {\em critical} value of
the binding constant $K^* =(1+\sqrt 2)/2$. At this special value the
amplitude $A_-$ vanishes and the function $G(\rho)$ becomes smooth and
monotonic in the physical regime $0<\rho<1/2$.

\begin{figure}[htbp]
  \begin{center}
     \psfrag{a}{$(a)$}
     \psfrag{b}{$(b)$}
    \psfrag{a*}{$\alpha=\alpha^*$}
    \psfrag{b*}{$\beta=\beta^*$}
    \psfrag{r}[][][1.2]{$\rho$}
    \psfrag{ri}{$\rho_I$}
    \psfrag{r*}{$\rho^*$}
    \psfrag{x}[][][1.2]{$x$}
    \psfrag{W-}{$W_{-1}$}
    \psfrag{W0+}{$W_{0}^{+}$}
    \psfrag{W0-}{$W_{0}^{-}$}
    \includegraphics[width=\columnwidth]{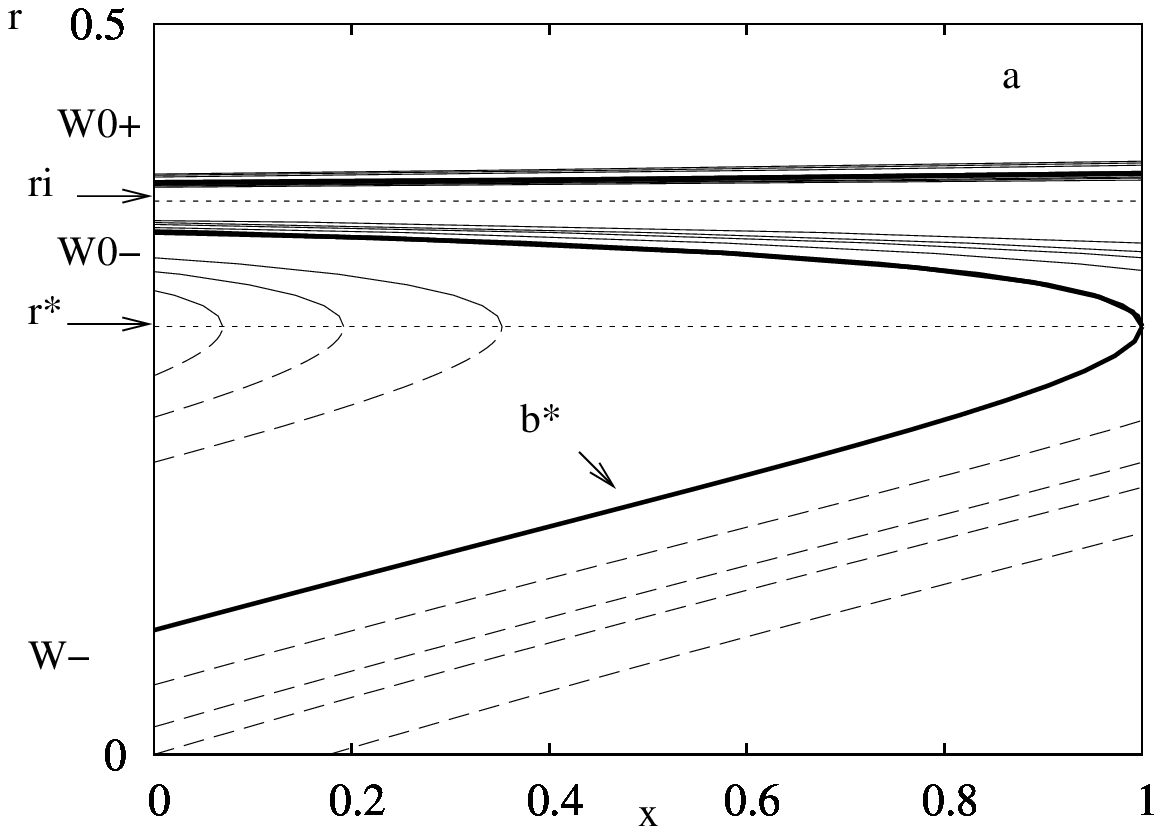}
    \includegraphics[width=\columnwidth]{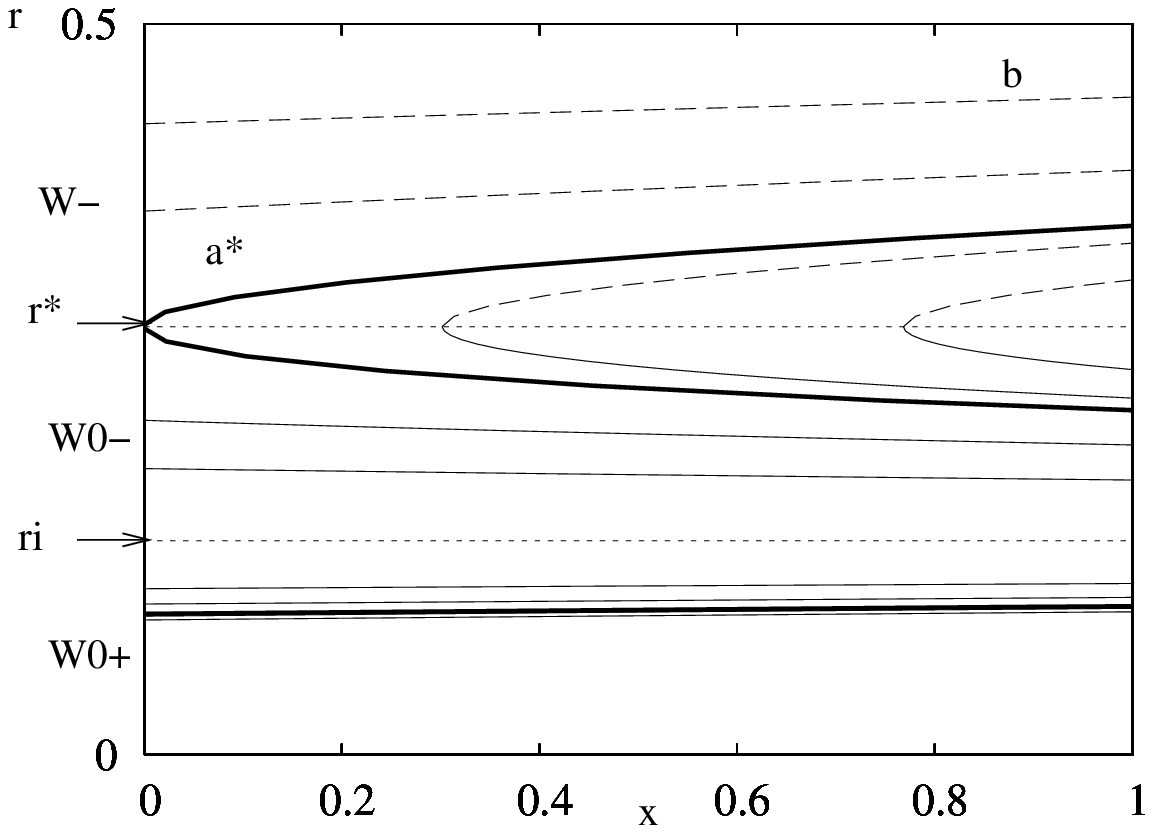}
    \caption{ \label{fig:fun} The functions $W_{-1}$ (dashed lines),
      $W_{0}^{-}$ (full lines), $W_{0}^{+}$ (full lines) for (a)
      $K=4$, $\Omega_D=0.05$ (high binding affinity) and (b) $K=0.25$,
      $\Omega_D=0.05$ (low binding affinity). The branches $W_{0}^{+}$ and
      $W_{0}^{-}$ are separated by the isotherm density $\rho_I$.
      Varying $\alpha$ or $\beta$ results in a simple shift parallel
      to the $x-$axis.  In Fig.(a/b) the thick lines represent the
      solutions where the branch point hits the right/left end of the
      system. The dotted lines indicate the critical density value
      $\rho^*$ (where branch point occurs) and the isotherm
      $\rho_{I}$.}
  \end{center}
\end{figure}

After matching the boundary conditions, the left and right solutions
are obtained up to inversion of a function
\begin{subequations}
  \begin{equation}
    \label{eq:fa}
    L_\alpha(\rho_\alpha)\equiv G(\rho)-G(\alpha/(1+\alpha))=x\, ,
  \end{equation}
  \begin{equation}
    \label{eq:fb}
    R_\beta(\rho_\beta)\equiv G(\rho)-G((1-\beta)/2)+1=x\, .
  \end{equation}
\end{subequations}
Upon inverting, the singularity of $G(\rho)$ transforms into a
horizontal asymptote, whereas the extremum manifests itself as a
branch point, see Fig.~\ref{fig:fun}. The functions
are multivalued with three branches, to be referred as $W_{0}^{+}$,
$W_{0}^{-}$ and $W_{-1}$.

\subsection{High binding affinity: $K>K^*$}
For large binding constants, $K>K^*$, the equilibrium density of the
on-off isotherm is larger than the critical density at which the current
of pure TASEP is maximal, $\rho_I>\rho^*$, see Fig.~\ref{fig:fun}.

For densities smaller than $\rho_I$, the track accumulates particles
from the reservoir, since
\begin{eqnarray}
\label{eq:signs}
F_A-F_D=\Omega_D\frac{1+4K}{1-\rho}\left[(\rho-\rho_-)(\rho-\rho_+)\right]\, .
\end{eqnarray}
Hence the isotherm acts as an attractor. This property has to be
reflected in the left and right solution solving the balance equation
for the current, $\partial_xj=(\partial_\rho
j)(\partial_x\rho)=F_A-F_D$.  For $\rho(0)=\alpha/(1+\alpha) >
\rho^*$, the branches $W_0^\pm$ are unstable. For densities $\rho>
\rho_I$ the coupling to the reservoir induces a decrease in density,
whereas $W_0^+$ is increasing.  Similarly for $\rho^* < \rho < \rho_I$
the net gain due to the on/off kinetics is not reflected by the
decreasing function $W_0^-$. One concludes that for the left solution
the only physical branch is $W_{-1}$. For
$\alpha>\alpha^*=\rho^*/(1-\rho^*)=1/(1+\sqrt 2)$, i.e.\ the critical
entrance rate of dimer TASEP, a left boundary layer necessarily
appears.

For the right solution the argument is just reversed, i.e.\ the
physical branches are $W_0^\pm$. The critical exit rate $\beta^*$ of
dimer TASEP is determined by $\rho(1)=(1-\beta^*)/2=\rho^*$ and equals
$\beta^*=1/(1+\sqrt 2)$. For exit rates larger than the critical one a
right boundary layer emerges.
\begin{figure}[htbp]
  \begin{center}
    \psfrag{beta}[][][1.2]{$\beta$} 
    \psfrag{alpha}[][c][1.2]{$\alpha$}
    \psfrag{b*}{$\beta^*$} 
    \psfrag{HD}[][][1]{HD}
    \psfrag{LD/HD}[][][1]{LD/HD}
    \psfrag{LD}[][][1]{LD}
    \psfrag{LD/M}[][][1]{LD/M}
    \psfrag{M}[][][1]{M}
    \psfrag{l}{$l$}
    \psfrag{r}{$r$}
    \psfrag{rl}{$\rho_\alpha$}
    \psfrag{rh}{$\rho_\beta$}    
    \psfrag{rr}{$\rho_\beta$}
    \psfrag{rlm}{$\rho_\alpha/\rho_{\beta^*}$}
    \psfrag{rlh}{ $\rho_\alpha/\rho_\beta$}
    \psfrag{rm}{ $\rho_{\beta^*}$}
    \psfrag{X0}{$x_w=0$}
    \psfrag{X1}{$x_w=1$}
    \includegraphics[width=\columnwidth]{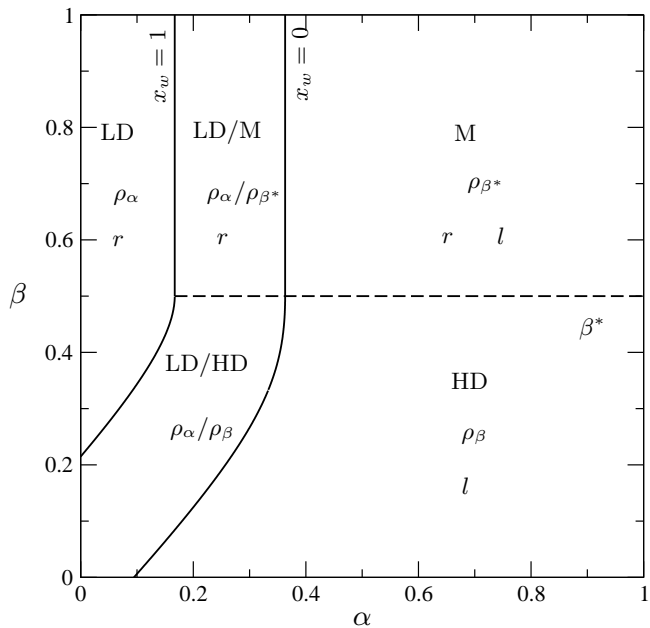}
    \caption{ \label{fig:pdk2}High binding affinity; phase diagram for
      $K=2$, $\Omega_D=0.1$. In the low-density (LD) phase the profile
      is given by the left solution $\rho_\alpha$ and a boundary layer
      appears at the right ($r$).  The position of boundary layers
      ($r$, $l$), as well as the relevant bulk solution
      ($\rho_\alpha$, $\rho_\beta$, and $\rho_{\beta^*}$) are
      indicated for all phases (HD: high density phase, LD/HD: low
      density-high density coexistence phase, M: Meissner phase, LD/M:
      low density-Meissner coexistence phase, see text).}
    \end{center}
\end{figure}
The complete phase diagram is obtained by identifying different
scenarios to match left and right solutions. The boundary conditions,
Eq.~(\ref{eq:bcpp}) can be satisfied by the stable branches of the
left and right solution only for $\alpha <\alpha^*, \beta<\beta^*$
respectively. The merging of the two solutions by
Eq.~(\ref{eq:condition}) is achieved at some position $x_w$. Depending
on whether the (i) $0<x_w<1$, (ii) $x_w<0$ and (iii) $x_w>1$, a domain
wall in the bulk emerges (i), a boundary layer at the left (ii) or
right (iii) appears. For $\beta > \beta^*$ an additional boundary
enters the system. The result of such an analysis is presented in
Fig.~\ref{fig:pdk2}, and the chain of arguments follows the one of
Ref.~\cite{parmeggiani-franosch-frey:04}.

For example, for large entrance rate $\alpha$ and low exit rates
$\beta$ the density profile is in a high density phase (HD), i.e.\ the
solution $\rho(x)$ exceeds $\rho^*$, and is given by the stable right
solution $\rho_\beta(x)$. At the left end $x=0$ a boundary layer
appears. Similarly a region of a low density phase (LD) and low
density-high density phase coexistence (LD/HD) with appropriate
boundary layer and domain wall is inferred from the figure.

For $\beta>\beta^*$ the right solution is replaced by the critical
right solution $\rho_{\beta=\beta^*}$, which constitutes the analogue
to the maximal current phase in TASEP: the system indeed cannot
transport more than the current $j_{\beta^*}$ imposed by the critical
exit rate $\beta^*$.

Changing the exit rate beyond $\beta^*$ does not affect the density
profile except in a small boundary layer. This phenomenon is similar
to a superconductor where an external magnetic field does not enter
the sample except for a short penetration depth.  This analogy
suggests to denote this region in the phase diagram by M (Meissner)
phase. The coexistence of low-density and Meissner phase (LD/M)
extends the coexistence phase (LD/HD) for $\beta > \beta^*$.

The phase boundaries can be computed analytically up to inversion of a
function. For example, the HD - LD/HD phase boundary is obtained by
requiring the domain wall to fall on the left end of the system $x_w =
0$. Thus the continuity condition, Eq.~(\ref{eq:condition}) translates the
boundary condition $\rho(0)= \alpha/(1+ \alpha)$ to $\rho(x + 0) =
(1-\alpha)/2$. In terms of the implicit solution, Eq.~(\ref{eq:fb}), this
implies
\begin{eqnarray}
\label{eq:dw0}
R_\beta\left(\frac{1-\alpha}{2}\right)=0 \, .
\end{eqnarray}
Similarly the LD- LD/HD phase boundary is determined by placing the
domain wall at $x_w = 1$, which yields
\begin{eqnarray}
\label{eq:dw1}
L_\alpha\left(\frac{\beta}{1+\beta}\right)=1 \, .
\end{eqnarray}
For $\beta > \beta^*$ the preceding equations still hold provided
$\beta$ is replaced by the critical exit rate $\beta^*$.\cite{foot6}
Correspondingly, phase boundaries degenerate to straight vertical
lines.

\begin{figure}[htbp]
  \begin{center}
    \psfrag{r}[][][1.2][-90]{$\rho^c$}
    \psfrag{x}[][][1.2]{$x$}
    \psfrag{(a)}{(a)}
    \psfrag{r1}{HD $(0.6,0.2)$}
    \psfrag{r2}{M $(0.6,0.6)$}
    \psfrag{r3}{HD/M $(0.2,0.6)$} 
    \psfrag{r4}{L $(0.1,0.6)$}     
    \vspace{0cm}
    \includegraphics[width=\columnwidth]{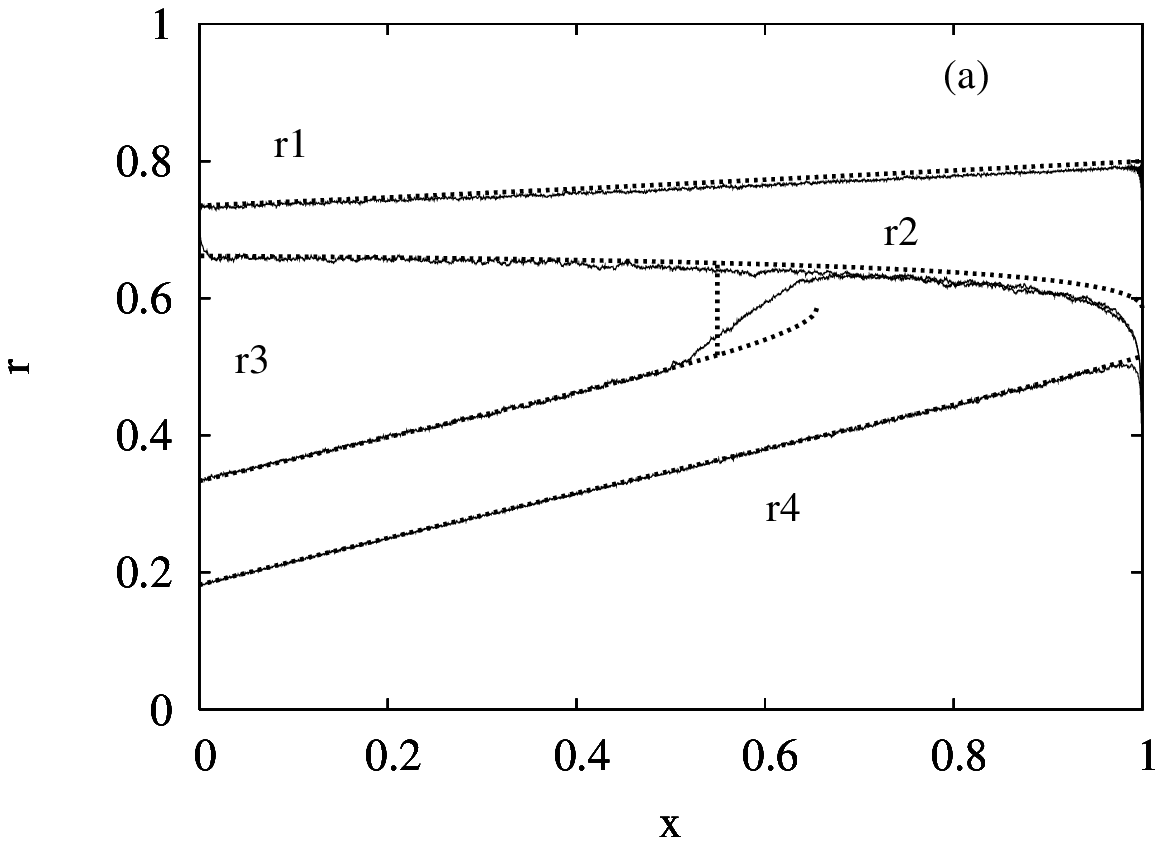}
    \psfrag{N}{$N$ increases}
    \psfrag{(b)}{(b)}
    \psfrag{r1}{$\rho_\beta$}
    \psfrag{r2}{$\rho_\alpha$}
    \includegraphics[width=\columnwidth]{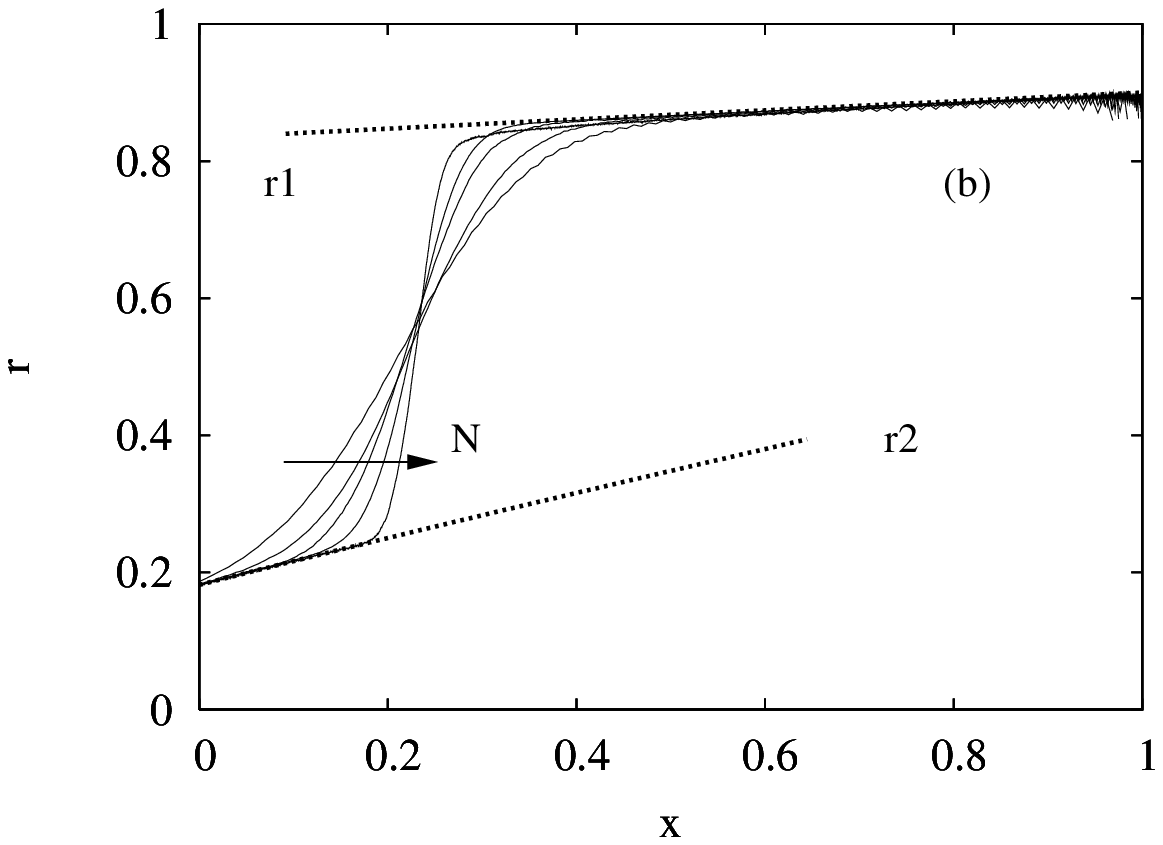}
    \caption{ \label{fig:example1}(a)~Density profiles for $K=2$ and
      $\Omega_D=0.1$ in systems of $N=4096$ lattice sites; parameters
      $(\alpha,\beta)=(0.1,0.6), (0.2,0.6),(0.6,0.6),(0.6,0.2)$.
      (b)~Approach of the mesoscopic limit of the density profile for
      $\alpha=\beta=0.1$, $\Omega_D=0.1$, $K=2$ and different system
      sizes ($N=128,256,512,1024,4096$); the domain wall becomes
      steeper with the increasing of $N$. Wiggly lines represent
      simulated data, while dotted lines indicate analytic solutions}
 \end{center}
\end{figure}

Density profiles exemplifying different regions of the phase diagram
are showed in Fig.~\ref{fig:example1} and compare nicely to the
corresponding simulated data.

\begin{figure}[htbp]
  \begin{center}
    \psfrag{K}[][][1]{$K$} 
    \psfrag{om}[][][1]{$\Omega_D$}
    \psfrag{K*}{$K^*$}     
    \psfrag{om1}{$\Omega_{D1}$}
    \psfrag{om2}{$\Omega_{D2}$}
    \psfrag{oma}{$\Omega_a^*$}
    \psfrag{omb}{$\Omega_b^*$}
    \psfrag{kin}{$\Omega_D^*\!\!\!\searrow\!0$}
    \vspace{0cm}
    \includegraphics[width=0.9\columnwidth]{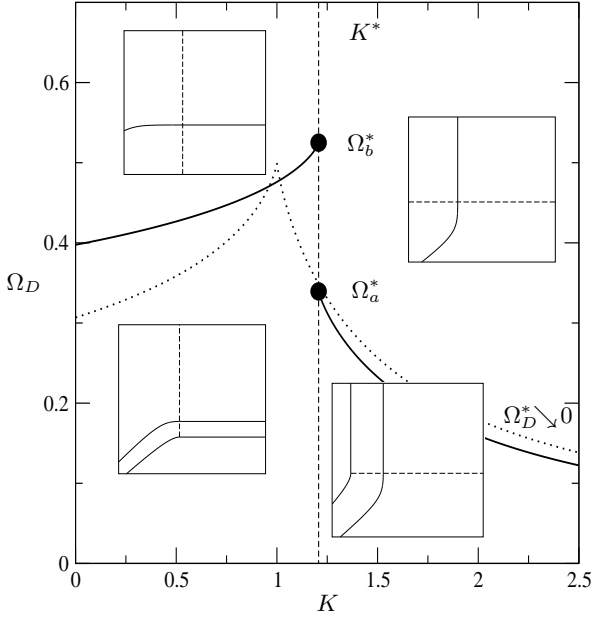}
    \caption{ \label{fig:komega}Critical detachment rate $\Omega_D^*$
      as a function of the binding constant $K$ (solid line shows
      $\Omega_D^*(K)$ for dimers, dotted line for monomers). The
      dashed line $K^*$ separates the cases of ``high'' form the
      ``low'' binding affinity (for dimers). For large $K$ the
      critical detachment rate approaches zero.  Insets represent the
      different topologies of the $\alpha-\beta$ section of the phase
      diagram.}
  \end{center} 
\end{figure}

The low density phase disappears from the phase diagram upon
increasing $\Omega_D$ at fixed $K$. Such a topological change occurs,
once the phase boundary of the LD/M coexistence phase hits the
$\beta$-axis of the phase diagram (i.e.\ $\alpha=0$). The condition is
inferred from Eq.~(\ref{eq:dw1}) and reads:
\begin{eqnarray}
  \label{eq:critdw1}
  L_{\alpha=0}\left(\frac{\beta^*}{1+\beta^*}\right)=1\, .
\end{eqnarray}
A numerical solution of the preceding equation for $\Omega_D^*$ as a
function of $K$, is presented in Fig.~\ref{fig:komega}.

In the limit $\Omega_D\to 0^+$ the LD/HD coexistence phase shrinks
continuously to the line $\alpha=\beta$ and one recovers the dimer
TASEP phase diagram. For $\Omega_D\gg\Omega_D^*$, the coexistence
phases constitute only a marginal region in the $\alpha-\beta$
plane, located close to the $\beta$-axis. The phase diagram is
dominated by the HD and M phases, however the density profile
approaches the constant value $\rho_I$ in the bulk, as expected from
pure on-off kinetics.\cite{foot7}

\subsection{Low binding affinity: $K<K^*$}
If $K<K^*$ the functional form of the solutions is still given by
Eqs.~(\ref{eq:integrate},\ref{eq:fa},\ref{eq:fb}). However, now the isotherm $\rho=\rho_I$ lies
below the critical density, $\rho_{I}<\rho^*$. By the same reasoning
as above, see Eq.~(\ref{eq:signs}), the left stable branches are the one
which approach the isotherm, $W_{0}^{\pm}$ in
Fig.~\ref{fig:fun}b, while the only stable solution for the
right side is $W_{-1}$, i.e.\ reversed to $K>K^*$.

Using the same methods as for the high binding constant, phases and
phase boundaries are identified by matching the appropriate left and
right solutions. A cut of the parameter space at fixed $\Omega_D$ and
$K$ is presented in Fig.~\ref{fig:komega}. The high density phase
disappears from the phase diagram for detachment rates exceeding a
critical value $\Omega_D^*(K)$, see
Fig.~\ref{fig:komega}.\cite{foot8}

Let us stress that there is an important difference with respect to
the monomeric analogue, namely the particle-hole symmetry is no longer
present.  Correspondingly, a symmetry transformation does not yield
the phase diagram of the ``low affinity'' from the one of the
``high affinity''. Yet, the exchange of high densities and low
densities, left and right, exit by entrance rates yields the correct
topology of the phase diagram as well, as qualitatively the density
profile.
\begin{figure}[htbp]
  \begin{center}
    \psfrag{beta}[][c][1.2]{$\beta$} 
    \psfrag{alpha}[][c][1.2]{$\alpha$}
    \psfrag{a*}{$\alpha^*$} 
    \psfrag{r}{$r$}
    \psfrag{l}{$l$}
    \psfrag{X0}{$x_w=0$}
    \psfrag{X1}{$x_w=1$}
    \psfrag{M/HD}[][][1]{M/HD}
    \psfrag{M}[][][1]{M}
    \psfrag{b*}{$\beta^*$} 
    \psfrag{HD}[][][1]{HD}
    \psfrag{LD/HD}[][][1]{LD/HD}
    \psfrag{LD}[][][1]{LD}
    \psfrag{LD/M}[][][1]{LD/M}
    \psfrag{M}[][][1]{M}
    \psfrag{l}{$l$}
    \psfrag{r}{$r$}
    \psfrag{rm}{$\rho_{\beta^*}$}    
    \psfrag{rl}{$\rho_\alpha$}
    \psfrag{rh}{$\rho_\beta$}    
    \psfrag{rmh}{$\rho_{\beta^*}/\rho_\beta$}
    \psfrag{rlh}[][]{$\rho_\alpha/\rho_\beta$}
    \vspace{1cm}
    \includegraphics[width=\columnwidth]{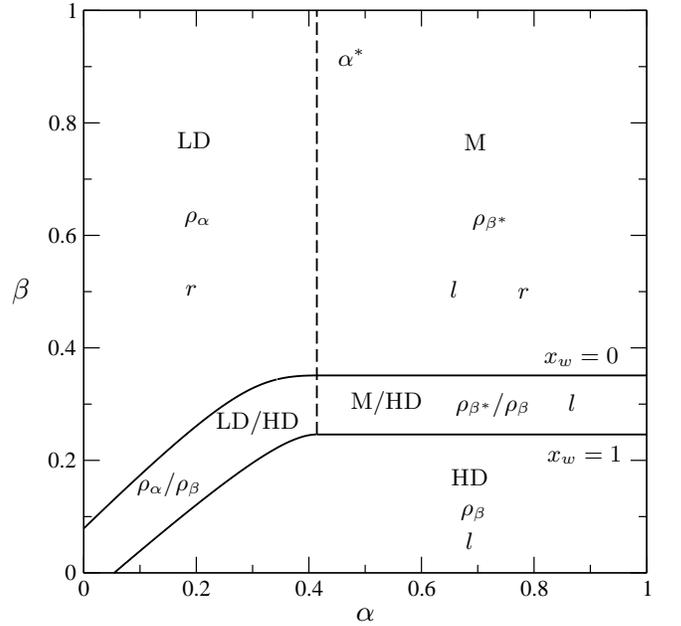}
    \caption{ \label{fig:pdk0.8}Low binding affinity; phase diagram
      for $K=0.8$, $\Omega_D=0.1$.  Legend as in
      Fig.~\ref{fig:pdk2}.}
    \end{center}
\end{figure}

\subsection{The critical case $K=K^*$}
\label{sec:criticalk}
The limiting case when the binding constant equals the critical one,
$K=K^*$, requires a separate analysis, since the amplitude $A_-$
vanishes, cancelling exactly the logarithmic divergence at
$\rho=\rho_-$ (\emph{nota bene} $\rho_-=\rho^*=\rho_I$). This signals
that $\rho=\rho^*=const$ is a spurious solution of the balance
equation for the stationary density profile. It corresponds to the
maximal current (MC) phase of pure dimer TASEP.  Indeed for $K=K^*$
Eq.~(\ref{eq:MFcont2}) simplifies to
\begin{eqnarray}
  \label{eq:ksimpl1}
  (\rho-\rho^*)\left[\frac{2\rho-(2+\sqrt{2})}{(1-\rho)}\partial_x\rho-\Omega_D\left(\rho-\frac{1}{\sqrt{2}}\right)\right]=0\, . \nonumber\\ 
\end{eqnarray}
The non-trivial solution Eq.~(\ref{eq:integrate}) specializes to
\begin{eqnarray}
  \label{eq:ksimpl2}
  G(\rho)=\frac{2\sqrt 2-2}{\Omega_D}\left(\ln(1-\rho)-\sqrt 2\ln\left(\frac {1}{\sqrt{2}}-\rho\right)\right)\, ,
\end{eqnarray}
which is a monotonic function in the physical regime. Correspondingly,
the inverse function is single valued. 

The properties of the phase diagram for $K=K^*$ are obtained by
assembling the resulting left and right branches, as well as the
constant critical density $\rho^*$, for different boundary conditions.
We have determined numerically the line where the domain wall hits the
left and right end, $x_{w}=0$ and $x_{w}=1$, respectively by solving
Eqs.~(\ref{eq:dw0} and \ref{eq:dw1}).  Moreover both left and right
solutions now can approach the critical one (i.e.\ $\rho=\rho_{I}$) in
the bulk and therefore a triple phase coexistence LD-MC-HD is
possible. This phase is bounded by the lines $\beta=\beta^*$ above and
$\alpha=\alpha^*$ on the right. Furthermore the curve
$g(\alpha,\beta)$, where the left and right currents match
additionally the critical one, separates the triple phase coexistence
from the LD-HD phase. Implicitly $g(\alpha,\beta)$ is given by the
condition
$L_\alpha\left(\beta^*/(1+\beta^*)\right)=R_\beta\left((1-\alpha^*)/2\right)$.

\begin{figure}[htbp]
  \begin{center}
    \psfrag{beta}[][][1.2]{$\beta$}
    \psfrag{alpha}[][][1.2]{$\alpha$}
    \psfrag{a*}{$\alpha^*$}
    \psfrag{b*}{$\beta^*$}
    \psfrag{g}{$g(\alpha,\beta)$}
    \psfrag{LD}{LD}
    \psfrag{HD}{HD}
    \psfrag{LD/HD}{LD/HD}
    \psfrag{MC/HD}{MC/HD}
    \psfrag{LD/MC}{LD/MC}
    \psfrag{MC}{MC}
    \psfrag{LD/MC/HD}{LD/MC/HD $\rho_\alpha/\rho^*/\rho_\beta$}
    \psfrag{l}{$l$}
    \psfrag{r}{$r$}
    \psfrag{rl}{$\rho_\alpha$}
    \psfrag{rh}{$\rho_\beta$}    
    \psfrag{rr}{$\rho_\beta$}
    \psfrag{rlm}{$\rho_\alpha/\rho^*$}
    \psfrag{rmh}{$\rho^*/\rho_\beta$}
    \psfrag{rlh}{ $\rho_\alpha/\rho_\beta$}
    \psfrag{rm}{ $\rho^*$}
    \psfrag{X0}{$x_w=0$}
    \psfrag{X1}{$x_w=1$}
    \vspace{15mm}
    \includegraphics[width=\columnwidth]{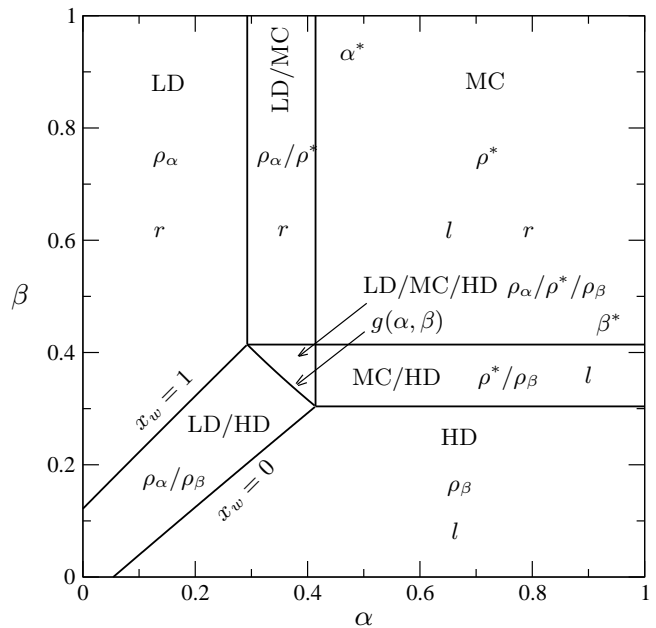}
    \caption{ \label{fig:pdkc}Critical case; phase diagram for the case
      $K=K^*$, $\Omega_D=0.1$.  Seven phases appear: LD, MC, HD and
      all the possible coexistence: LD/MC, MC/HD, LD/HD, LD/MC/HD.
      Legend as in Fig.~\ref{fig:pdk2}.}
  \end{center}
\end{figure}

\begin{figure*}[htbp]
 \begin{center}
 \psfrag{beta}[][][1.2]{$\beta$}
 \psfrag{LD}{LD} \psfrag{HD}{HD} \psfrag{LD/HD}{LD/HD}
 \psfrag{MC/HD}{MC/HD} \psfrag{LD/MC}{LD/MC} \psfrag{MC}{MC}
 \psfrag{LD/MC/HD}{LD/MC/HD} \psfrag{0}{} \psfrag{1}{}
 \psfrag{(a)}{(a)} \psfrag{(b)}{(b)} \psfrag{(c)}{(c)}
 \psfrag{a}{$\alpha$} \psfrag{b}{$\beta$}
 \begin{tabular}{lcc}
       \subfigure{\includegraphics[width=0.61\columnwidth]{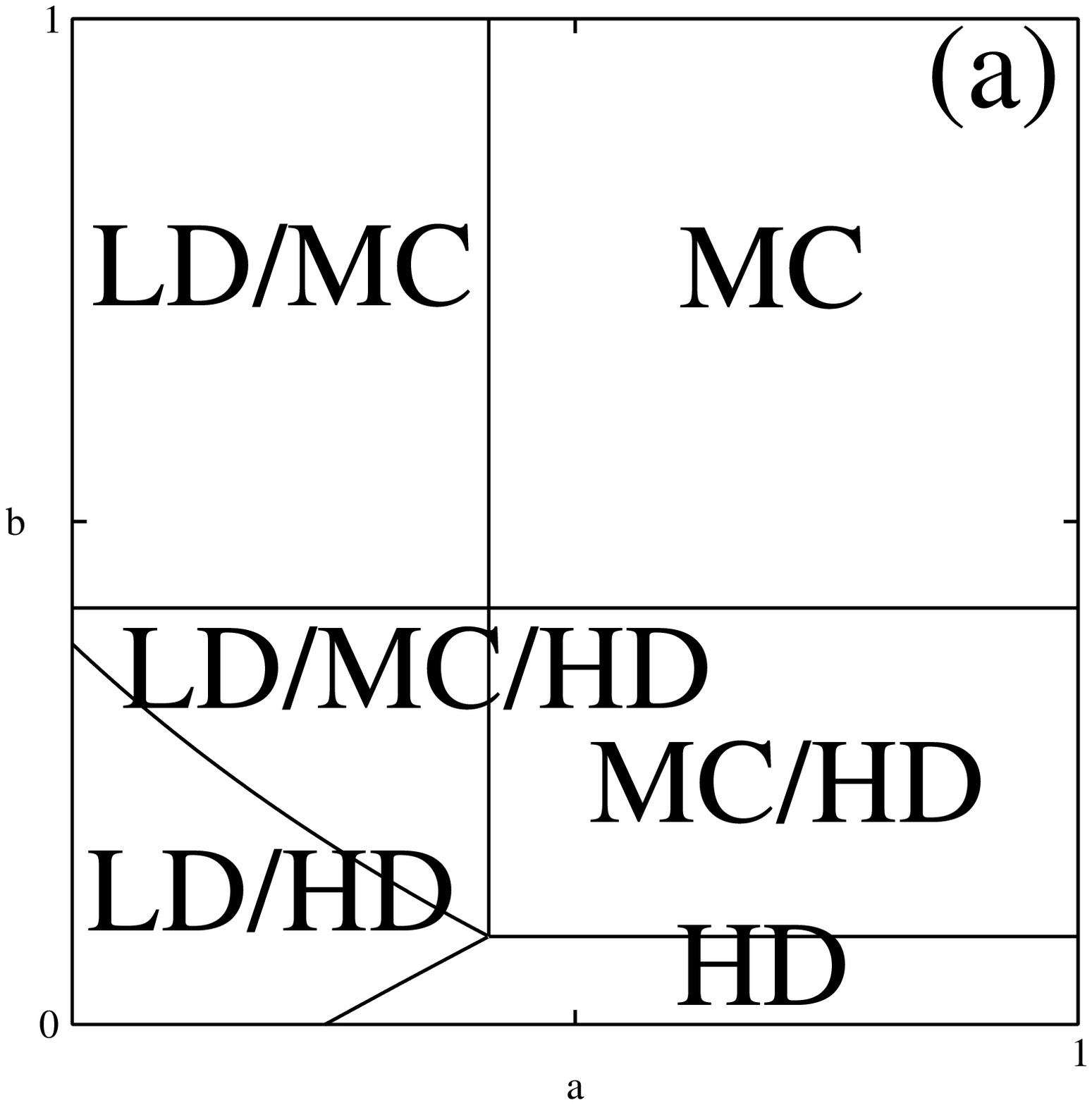}}&
       \hspace{8mm}\subfigure{\includegraphics[width=0.61\columnwidth]{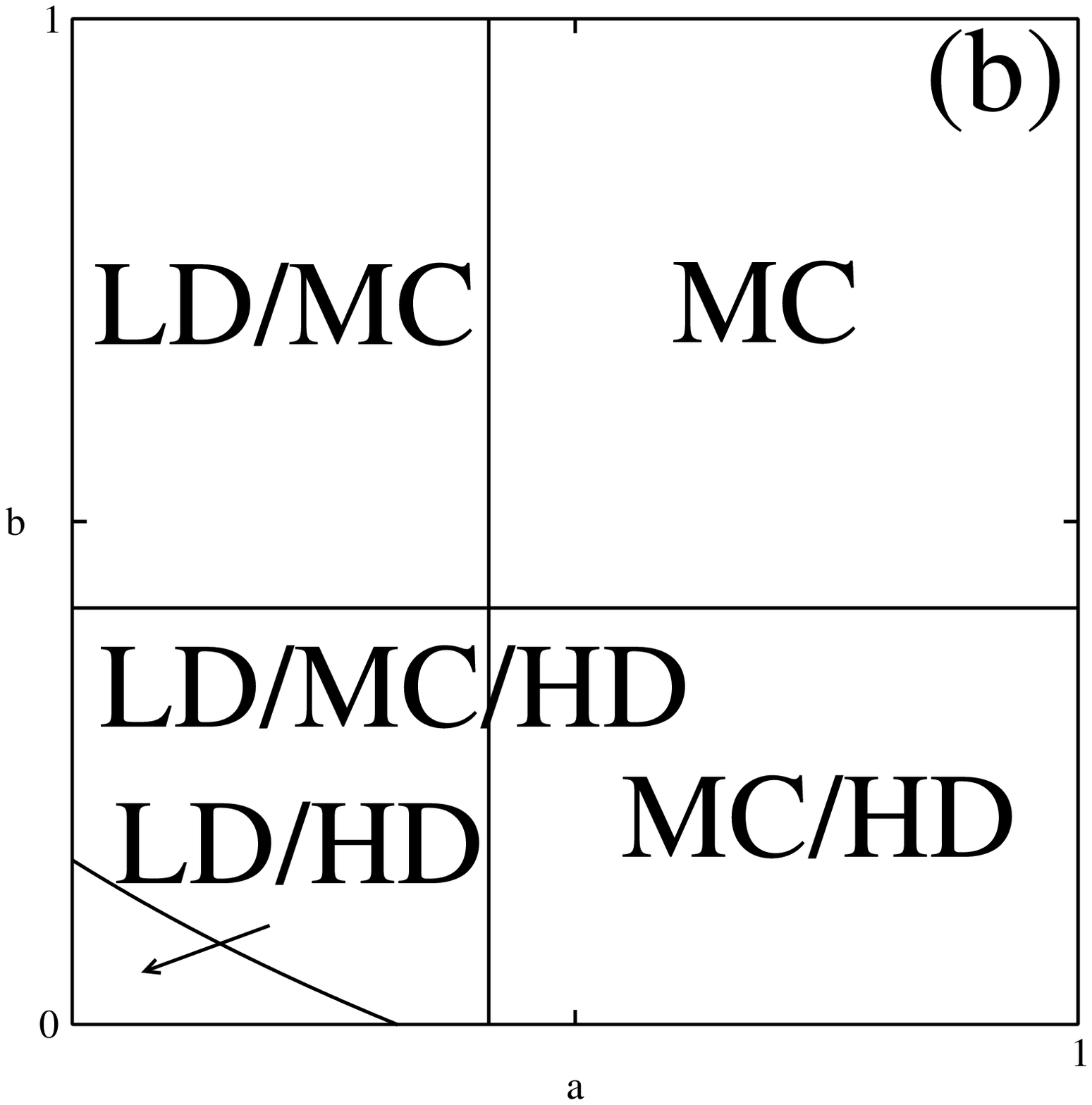}}&
       \hspace{8mm}\subfigure{\includegraphics[width=0.61\columnwidth]{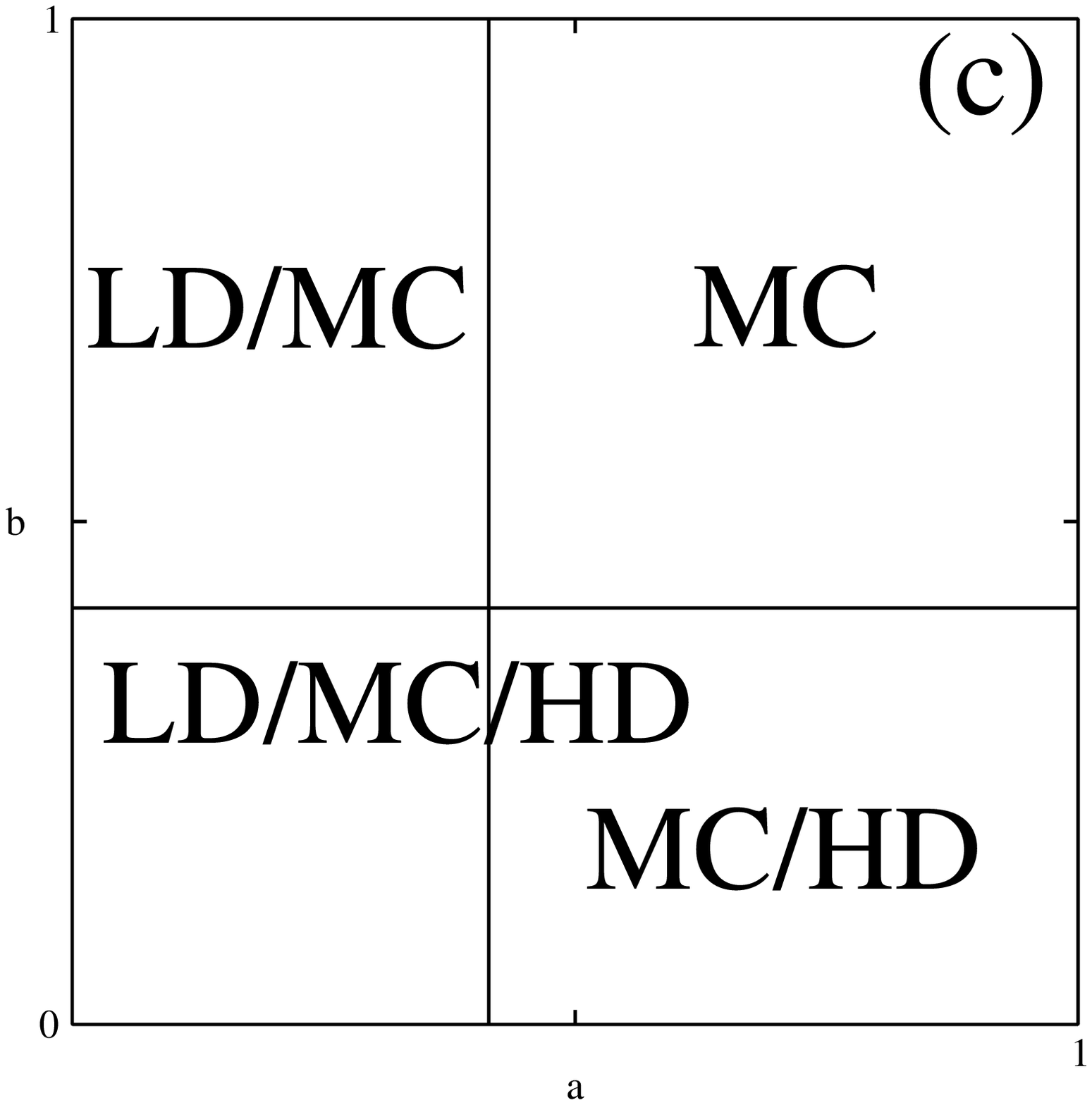}}      
     \end{tabular}
      \caption{ \label{fig:kctopol}Different topologies of the phase
        diagram for $K=K^*$ and (a) $\Omega_a^*<\Omega_D<\Omega_b$,
        (b) $\Omega_b^*<\Omega_D<\Omega_c^*$ and (c)
        $\Omega_D>\Omega_c^*$.  Phases are indicated in the graphs.}
   \end{center}
 \end{figure*}

Depending on the value of the kinetic rate $\Omega_D$, different
topologies for the phase diagram arise. For example, for low
$\Omega_D$ seven phases are present in the $\alpha-\beta$ diagram (see
Fig.~\ref{fig:pdkc}. Upon increasing $\Omega_D$, a first topological
change, analogous to the case of high binding affinity, occurs, i.e.\
the LD phase exits the phase diagram. By solving
Eq.~(\ref{eq:critdw1}), one obtains the critical value
$\Omega_a^*\approx 0.33945$, see Fig.~\ref{fig:komega}.  At another
critical value $\Omega_b^*\approx 0.52496$ the HD phase disappears,
and at still higher detachment rates $\Omega_D>\Omega_c^*=0.86441$ the
LD-HD coexistence phase is no more present and the phase diagram (but
not the density profiles) is independent of $\Omega_D$. Representative
graphs of these topologies are shown in Figs.~\ref{fig:komega}
and \ref{fig:kctopol}.

\section{Conclusions}
\label{sec:conclusion}
We have studied the non-equilibrium dynamics of a 1D driven lattice
gas with open boundaries, where individual particles are considered as
dimers occupying two lattice sites simultaneously.  Upon combining a
generalized mean-field theory with extended stochastic simulations, we
have explored the phase behavior of the non-equilibrium steady state
as a function of the kinetic rates.  In particular, we have analyzed
the interplay between the coupling of the system to a reservoir at the
boundaries and at the bulk.

In the following we compare our results with those obtained for the
analogous problem of simple TASEP of monomers coupled to Langmuir
kinetics. The balance equation in the stationary state that
constitutes the equivalent of Eq.~(\ref{eq:MFcont}) has been derived
in \cite{parmeggiani-franosch-frey:03,parmeggiani-franosch-frey:04}:
\begin{eqnarray}
  \label{eq:pff}
  \frac {1}{\Omega_D}\partial_x \left[\rho(1-\rho)\right]=K(1-\rho)-\rho\, .
\end{eqnarray}
Although the structure of the equation is the same, $\partial_x
j=F_A-F_D$, the current and the source terms obey a rather different
functional dependence.  In particular, the equation exhibits a
particle-hole symmetry, i.e.\ density profiles and the complete phase
diagram is symmetric with respect to the following transformation rule
$\rho(x)\leftrightarrow 1-\rho(1-x)$, $\Omega_D\leftrightarrow
\Omega_A$, $\alpha\leftrightarrow \beta$. Also, a TASEP for dimers
disregarding the on-off kinetics can be cast in a particle-hole
symmetric form, introducing an effective hole density $\overline\rho$
by the requirement $j(\rho)=j(\overline\rho)$ and interchanging
$\alpha$ with $\beta$; this explain why the phase diagram is symmetric
along the diagonal.  Similarly, a pure on/off kinetics is invariant
with respect to the transform $\Omega_A\leftrightarrow\Omega_D$ and an
effective hole density $\overline\rho=(1-2\rho)^2/(1-\rho)$. However,
one may easily show that the two symmetries of simple dimer TASEP and
on/off kinetics never coincide for any $K>0$, implying that the
coupling of both processes breaks the particle-hole symmetry.

It is remarkable that the topology of the resulting phase diagram is
almost unaffected by this broken symmetry. In fact, the single
qualitative change of topology occurs at $K=K^*$ for intermediate
$\Omega_D$, see Figs.~\ref{fig:komega} and \ref{fig:kctopol}b. The
origin of the \emph{robustness} of the picture can be traced back to
the following observations: (i) in a mean-field approach the balance
equation for the stationary density profile is an autonomous first
order ordinary differential equation; (ii) the current-density
relation exhibits a single extremum for both cases \cite{foot9}; (iii)
the solution $\rho(x)$ is obtained by quadrature and inversion of a
function resulting in a multivalued function with a simple branch
point (due to the maximum in the current-density relation) and a
horizontal asymptote (due to the on/off isotherm). Although the shape
of the explicit solution depends on the details of the density current
relation and the functional form of the on/off kinetics, the overall
topological features remain unchanged.  This reasoning explains that
particle hole symmetry is relevant for the topology of the phase
diagram only if the branching point is canceled by a zero of the
on/off kinetics, i.e.\ precisely at $K=K^*$.  We conclude that the
scenario for $K=K^*$ presented in Fig.~\ref{fig:kctopol} generalizes
the monomer TASEP coupled to Langmuir kinetics. Nevertheless there are
many differences in the numerical values of several quantities, which
are summarized in Table \ref{tab:diff}.

An interesting difference between monomers and dimers resulting from
the broken particle-hole symmetry can be read off from
Fig.~\ref{fig:komega}, where the change of the phase diagram topology
is shown as a function of $\Omega_D$ and $K$.  For monomers not only
the critical value $K^*$ is shifted, but the line $\Omega_D(K)$ is
continuous. Both the lines $\Omega_D=\Omega_D^*$ and $K=K^*$ mark
continuous transitions, but they concern two different events: while
at $K=K^*$ the shape of the analytical solution changes, at
$\Omega_D=\Omega_D^*$ a domain wall enters the system.  In the
monomers the critical detachment rates $\Omega_a^*$ and $\Omega_b^*$
coincide and the topology of the phase diagram in
Fig.~\ref{fig:komega} is such that it is impossible to go from a
configuration as in the bottom-left to the one in the top-right by
changing $K$ without crossing the transition line $\Omega_D^*$. The
(continuous) transition occurring at $\Omega_D=\Omega_D^*$ can be
studied experimentally since the density profile are remarkably
different. 

It is possible to generalize the continuous equation to the case of a
system containing $\ell$-mers, using directly the continuous results
presented in \cite{mcghee-hippel:74, shaw-zia-lee:03}:
\begin{eqnarray}
\label{lmers}
\frac{1}{\Omega_D}\partial_x\left[\frac{\rho(1-\ell\rho)}{1-(\ell-1)\rho}\right]=K\frac{(1-\ell\rho)^\ell}{(1-(\ell-1)\rho)^{\ell-1}}-\rho\, .\nonumber\\
\end{eqnarray}
The qualitative picture is contained already in the case of dimers;
although the functional form of the solution depends on the size of
the $\ell$-mers, the scenario of three branches, characterized by the
maximum of the current-density relation $\rho^*$ and the horizontal
asymptote $\rho_I$, is robust.  

\begin{table}[htbp]
\begin{center}
\begin{tabular}{|c||c|c|}
\hline
Quantity        & Monomers           & Dimers\\
\hline\hline
Particle dens.  & $\rho$        & $\rho$\\
\hline
Coverage dens.  & $\rho$        & $2\rho$\\
\hline
$j$             & $\rho(1-\rho)$        & $\frac{\rho(1-2\rho)}{1-\rho}$\\
\hline
$F_A$           & $\Omega_A(1-\rho)$    &
$\Omega_A\frac{(1-2\rho)^2}{1-\rho}$\\
\hline
$F_D$           & $\Omega_D\rho$    & $\Omega_D\rho$\\
\hline
$\rho(0)$   & $\alpha$    &$\frac{\alpha}{1+\alpha}$\\
\hline
$\rho (1)$   & $1-\beta$    &$\frac{1-\beta}{2}$\\
\hline
$\rho^*$        &$\frac 1 2$    &$\frac{1}{\sqrt{2}(1+\sqrt{2})}$\\
\hline
$\alpha^*=\beta^*$        &$\frac 1 2$    &$\frac{1}{1+\sqrt{2}}$\\
\hline
$j^*$        &$\frac 1 4$    &$\frac{1}{(1+\sqrt{2})^2}$\\
\hline
$j_{\alpha}$        &$\alpha(1-\alpha)$    &$\frac{\alpha(1-\alpha)}{1+\alpha}$\\
\hline
$j_{\beta}$        &$\beta(1-\beta)$    &$\frac{\beta(1-\beta)}{1+\beta}$\\
\hline
$\rho_I$ &$\frac{K}{1+K}$ &$\frac 1 2 \left(1-\frac{1}{\sqrt{1+4K}}\right) $\\
\hline
$K^*$        &$1$    &$\frac{1}{2}(1+\sqrt{2})$\\
\hline
\end{tabular}
\end{center}
\caption{
  \label{tab:diff}Differences and analogies between the model with monomers and 
  dimers: the current-density relation present in both cases a maximum 
  at $\rho^*$ which appears at different values,  but the particle-hole symmetry  which
  is trivial in the monomers is  actually effective in the dimers; the
  boundary conditions, as well as the current, need to be set   differently in the dimers in order to keep track of their geometry; as a consequence of this the critical values of the boundary rates are different from the
  critical density in the case of dimers; the the isotherm is related to the binding  constant in a non trivial way in the
  case of dimers; the particle-hole symmetry is preserved by the sources only in the
  monomers case. See the text for more details.   
}
\end{table}

In conclusion we have studied a driven one dimensional lattice gas of
dimers where the dynamics of the totally asymmetric exclusion process
has been coupled to the on-off kinetics in the bulk. We used the
ad-hoc ``refined'' mean-field for the TASEP part introduced in
previous works
\cite{macdonald-gibbs:69,lakatos-chou:03,shaw-zia-lee:03} and proved
that it is consistent with the assumptions made for the on-off part.
The main effect of extended nature of dimers on the phase behavior of
the system is related to the breaking of a symmetry in the problem
(particle-hole). This does have quantitative but not qualitative
consequences on the density profile and on the phase diagram.  The
origin of the \emph{robustness} of the picture found for monomers can
be traced back to the form of the stationary density profile which
depends exclusively on the form of the current-density relation and of
the isotherm of the on-off kinetics.  This suggests that the TASEP
dynamics washes out the interesting two-step relaxation dynamics that
characterizes the on-off kinetics of dimers. The non-trivial outcome
is that, in these systems, the diffusion (yet asymmetric) always
dominates the large time-scale relaxation. Further studies on the
dynamical correlation functions could point out more subtle
differences between the dynamics of dimers and monomers.  We
conjecture that quite interesting phenomena would arise upon taking
into account other interactions between the particles, which suggest
to consider alternative current-density relations (with more maxima)
or non-trivial on-off isotherm (cooperative on-off kinetics).

\acknowledgments We have benefited from discussions with Jaime Santos,
Andrea Parmeggiani, Mauro Mobilia, Hauke Hinsch and Tobias
Reichenbach.

\appendix 

\section{The conditional probability}
\label{app:cp}
In this appendix we derive an approximate expression for the joint
probability $p(i,0;i-1,0)$ and $p(i,2;i+1,0)$
starting from the mean-field assumption
\begin{eqnarray}
\label{eq:MFapprox1}
p(i,0|i-1,0)\simeq p(i,0|(i-1,0)\lor (i-1,2))\, .
\end{eqnarray}
Since the event $(i,0;i-1,1)$ is excluded by the dimer
geometry, the mean-field approximation assumes the occupancy of site
$i$ to be independent of the state of the previous site. Using Bayes
theorem, the r.h.s. of Eq.~(\ref{eq:MFapprox1}) can be expressed as
\begin{eqnarray}
  \lefteqn{ p((i,0)|(i-1,2) \lor (i-1,0))=}\nonumber\\
  & &\frac{p((i-1,0) \lor (i-1,2)| (i,0)) p(i,0)}{p((i-1,0) \lor
    (i-1,2))}\, .
\end{eqnarray}
The conditional probability in the numerator is unity, $p((i-1,0)\lor
(i-1,2)| (i,0))=1$, since a trail head cannot occupy site $i-1$ if
site $i$ is empty.  Moreover the events $(i,0)$ and $(i,2)$ in the
denominator are mutually exclusive $p((i-1,0) \lor
(i-1,2))=p(i-1,0)+p(i-1,2)$. Collecting results and using
Eq.~(\ref{eq:MFapprox1})
\begin{eqnarray}
\label{eq:condprob1}
  p(i,0|i-1,0)=\frac{p(i,0)}{p(i-1,0)+p(i-1,2)}\, .
\end{eqnarray}
The normalization condition compatible with the geometry of the dimers
allows to rewrite the denominator,
$p(i-1,0)+p(i-1,2)=1-p(i-1,1)=1-p(i,2)$, and one obtains Eq.~(\ref{eq:MF1}).

A similar argument can be used to compute the joint probability
$p(i,2;i+1,0)$, using $p(i+1,0|i,2)$ under the same initial
hypothesis
\begin{eqnarray}
\label{eq:MFapprox2}
p(i+1,0|i,2)\simeq p(i+1,0|(i,0)\lor(i,2))\, .
\end{eqnarray}
Using this equation and the normalization condition $p(i+1,0|i,0)
p(i,0)+ p(i+1,0|i,1) p(i,1)+p(i+1,0|i,2) p(i,2)=p(i+1,0)$ (together
with the fact that $p(i+1,0|i,1)=0$ for obvious geometric reasons),
one can show that $p(i+1,0|i,2)\simeq p(i+1,0|i,0)$. 

\section{Boundary conditions}
\label{app:bc}
We recall the arguments given previously
\cite{macdonald-gibbs:69,lakatos-chou:03,shaw-zia-lee:03} to justify
the boundary conditions Eq.~(\ref{eq:bcpp}).

\paragraph*{Left boundary:} 
The left boundary conditions in the continuum limit can be derived
from the rate equation. The incoming flux is given by the entrance
rate multiplied by the probability of having the first two sites free,
which is derived from the normalization condition on site $i=2$ (given
that site $i=1$ can never contain a trail head):
\begin{eqnarray}
  \label{eq:BCnorm}
  p(1,0;2,0)=1-p(2,2)-p(2,1)=1-\rho_{2}-\rho_{3}\, ,
\end{eqnarray}
while the outgoing flux is determined by the usual relation. This
yields to:
\begin{eqnarray}
  \label{eq:bcl1}
  \partial_{\tau}\rho_1&=&\alpha(1-\rho_{2}-\rho_{3})-\frac{\rho_{2}(1-\rho_{3}-\rho_{4})}{1-\rho_{3}}\, .
\end{eqnarray}
In the continuum limit (in first approximation) of the stationary
state $\rho_{2}=\rho_{3}=\rho$ and therefore the equation can be solve
for $\rho$ in terms of $\alpha$ to get Eq.~(\ref{eq:bcpp1})
\begin{eqnarray}
  \label{eq:bcl2}
  \rho(0)=\frac{\alpha}{1+\alpha}\, .
\end{eqnarray}
This can be seen as if the left density was imposed by the reservoir
density to which the system is coupled, which is not $\rho$, but
$\rho/(1-\rho)$, therefore
$\frac{\rho}{\rho_s}=\frac{\rho}{1-\rho}=\alpha$.  Note that the
current-density relation leads to the following boundary condition on
the current:
\begin{eqnarray}
  \label{eq:jBCL}
  j(0)=j_\alpha=\frac{\alpha(1-\alpha)}{1+\alpha}\, .
\end{eqnarray}

\paragraph*{Right boundary:} 
On the right boundary the last two sites are emptied at the same time
and there is no exclusion on the last site $i=N$, therefore the
incoming current on the second last site $i=N-2$ is the usual one
while the outgoing current is given by the occupation of the site
$i=N-1$ (since the last site is empty); consistently with the
continuity property, the outgoing current at site $i=N-2$ is also the
gain term for the last site, while the loss term is given by
$\beta\rho_N$:
\begin{subequations}
  \begin{equation}
  \label{eq:bcr1}
  \partial_{\tau}\rho_{N-1}=\frac{\rho_{N-2}(1-\rho_{N-1}-\rho_{N})}{1-\rho_{N-1}}-\rho_{N-1}\, ,
\end{equation}
\begin{equation}
  \label{eq:bcr2}
  \partial_{\tau}\rho_N=\rho_{N-1}-\beta\rho_N\, .
  \end{equation}
\end{subequations}
It is obvious that in the continuum limit the density in the last two
sites is different and only a coarse grained quantity like the average
density (which is half the coverage density) makes sense:
$\rho=(\rho_{N-1}+\rho_{N})/2$. One considers this last definition,
the stationary state of Eq.~(\ref{eq:bcr1}) with constant incoming
current (that gives $\rho(1-2\rho)/(1-\rho)=\rho_{N-1}$) and the
stationary state of Eq.~(\ref{eq:bcr2}) (that gives
$\rho_{N-1}=\beta\rho_N$), to obtain
\begin{eqnarray}
  \label{eq:bcr3}
  \frac{\rho(1-2\rho)}{1-\rho}=\rho\frac{2\beta}{1-\beta}\, ,
\end{eqnarray}
which finally gives the right boundary condition:
\begin{eqnarray}
  \label{eq:bcr4}
  \rho(1)=\frac{1-\beta}{2}\, .
\end{eqnarray}
The right boundary condition for the current reads:
\begin{eqnarray}
  \label{eq:jBCR}
  j(1)=j_\beta=\frac{\beta(1-\beta)}{1+\beta}\, ,
\end{eqnarray}
which has naturally the same form of the left condition because of the
effective particle hole symmetry.

Note also that in the HD phase the particles wait a long time before
detach giving rise to peculiar correlations: the (coverage) density
profile exhibits a sawtooth profile superimposed to the analytical one.
By using linear analysis on the discrete map it can be shown that
these correlations decay exponentially (but not as fast as the usual
boundary layers) to the fix point.


\begin{thebibliography}{40}
\bibitem{foot2}{A slight modification of the dynamical rules at the
    boundaries, e.g.\ allowing particles to enter (leave) the system
    gradually, has been shown not to introduce new features (see
    \cite{lakatos-chou:03,shaw-zia-lee:03} for discussion).}
\bibitem{foot4}{This approximation has been introduced first in the
    context of pure $\ell$-TASEP
    \cite{macdonald-gibbs-pipkin:68,macdonald-gibbs:69}.}
\bibitem{foot5}{The grand canonical partition function could be
    express in term of Chebyshev polynomial of the second kind
    $\mathcal{Z}_N=(-x)^{N/2}U_N\left(\frac{1}{2\sqrt{-x}}\right)$;
    equation~(\ref{eq:GCparttonks}) is merely a different
    representation.}
\bibitem{foot6}{For Eq.~(\ref{eq:dw1}) the replacement is obvious in
    the condition, whereas for Eq.~(\ref{eq:dw0}) notice that
    $R_\beta(\rho_\beta)$ depends on $\beta$ according to
    Eq.~(\ref{eq:fb}).}
\bibitem{foot7}{All the phase boundaries between LD-HD and HD phase
    intersect at a fix nodal point. In this point the
    solution must be independent of $\Omega_D$ but must depends on the
    binding constant and on the boundaries: this is the point where
    the bulk density imposed by the right boundary becomes equal to
    the isotherm, moreover since it is valid also in the limit of
    small $\Omega_D$ one expects $\alpha=\beta$. The coordinates of
    such a point $(\alpha_N,\beta_N)$ can be computed from
    $\rho_\beta(1)=(1-\beta)/2=\rho_{I}$ and are equal to
    $\alpha=\beta=1/\sqrt{4K+1}$.}
\bibitem{foot8}{In this case too there is a nodal point where all the
    phase boundaries between coexistence an HD (or M) phase (that
    indicate the domain wall at the left boundary) intersect: it can
    be computed from the condition
    $\rho_\alpha(0)=\alpha/(1+\alpha)=\rho_{I}$ and corresponds to the
    point $\alpha=\beta=(\sqrt{4K+1}-1)/(\sqrt{4K+1}+1)$.}
\bibitem{foot9}{This is not always the case: systems where
    interactions between particles are taken into account, as in
    \cite{hager-etal:01,popkov-etal:03} showed current-density
    relation with multiple maxima.}
\end{thebibliography}
\end{document}